\documentclass[12pt]{article}
\usepackage{jheppub}
\pdfoutput=1
\usepackage{amsmath,bbm,array,amsfonts,graphicx,wrapfig,lscape,float,mathtools,multirow,longtable,amsthm}
\usepackage[dvipsnames]{xcolor}
\usepackage{stackrel}
\usepackage[all]{xy}
\usepackage{graphicx}
\usepackage{caption}
\usepackage{subcaption}
\usepackage[bottom]{footmisc}
\usepackage{multirow}
\usepackage{mathrsfs}
\usepackage{tikz}
\usepackage{bm}
\usepackage{color}
\usepackage{xcolor}
\usepackage{enumerate}
\usetikzlibrary{decorations.pathreplacing}
\usepackage{diagbox}
\numberwithin{equation}{section}
\numberwithin{figure}{section}
\numberwithin{table}{section}
\usepackage{float}
\usepackage{pgfplots}
\pgfplotsset{compat=1.14}
\usepackage{longtable}
\usepackage{makecell}
\usepackage{lscape}
\usepackage[utf8]{inputenc}

\usepackage[autostyle=true]{csquotes}
\newtheorem{definition}{Definition}[section]
\newtheorem{theorem}{Theorem}[section]

\newtheorem{conjecture}[theorem]{Conjecture}

\usepackage[toc,page]{appendix}

\usepackage{tocloft}

\title{Chiral Rings, Futaki Invariants, Plethystics, and Gr\"obner Bases}

	\author[b]{Jiakang Bao,}
	\author[a,b,c]{Yang-Hui He,}
	\author[d]{Yan Xiao}

	\affiliation[a]{
		School of Physics, NanKai University, Tianjin, 300071, P.R. China}	
	\affiliation[b]{
		Department of Mathematics, City, University of London, EC1V 0HB, UK}
	\affiliation[c]{
		Merton College, University of Oxford, OX1 4JD, UK}
	\affiliation[d]{
		Department of Physics, Tsinghua University, Beijing, 100084, China}
	
	\emailAdd{jiakang.bao@city.ac.uk}
	\emailAdd{hey@maths.ox.ac.uk}
	\emailAdd{steven1025xiao@gmail.com}
	
	\preprint{
		\begin{flushright}
			
		\end{flushright}
	}

	\abstract{We study chiral rings of 4d $\mathcal{N}=1$ supersymmetric gauge theories via the notion of K-stability. We show that when using Hilbert series to perform the computations of Futaki invariants, it is not enough to only include the test symmetry information in the former's denominator. We discuss a way to modify the numerator so that K-stability can be correctly determined, and a rescaling method is also applied to simplify the calculations involving test configurations. All of these are illustrated with a host of examples, by considering vacuum moduli spaces of various theories. Using Gr\"obner basis and plethystic techniques, many non-complete intersections can also be addressed, thus expanding the list of known theories in the literature. 
	}

\begin{document}
	\maketitle

\newpage

\section{Introduction}\label{intro}
For supersymmetric gauge theories in 4d with $\mathcal{N}=1$, the chiral rings are important in the study of their dynamics;
this is the set of operators annihilated by  $\widehat{Q_{\dot{\alpha}}}$, defined modulo 
$\{ \widehat{Q_{\dot{\alpha}}}, \_ \}$, closed under addition and multiplication, whereby forming a ring structure.
In \cite{Collins:2016icw}, the interesting question of {\it when a polynomial ring is the chiral ring of a superconformal field theory} (SCFT) was posed. 
Since many new symmetries might emerge when a theory flows to IR (e.g. some free operators in the IR have these new symmetries acting on them), the idea of chiral ring \emph{stability} was introduced in \cite{Collins:2016icw} to determine whether there could be some new ring that would destabilize the original ring in the sense that the destabilizing ring would have a larger symmetry and would give no less central charge compared to the original ring\footnote{Notice that this does not violate the $a$-theorem which requires the central charge to decrease under RG flow since the original ring is not a ring of an SCFT.}. It was argued in \cite{Collins:2016icw} that this is equivalent to the concept of \emph{K-stability}\footnote{Therefore, we will use the words ``stability'' and ``K-stability'' interchangeably throughout.}. In \cite{Collins:2012dh,2015arXiv151207213C}, for a polarized ring with symmetry/Reeb vector field $\zeta$, K-stability is determined via perturbing the ring by a test symmetry $\epsilon \eta$ for some symmetry $\eta$ and small $\epsilon$.

The \emph{(Donaldson-)Futaki invariant}, which constitutes the criterion for K-stability, was originally defined in \cite{Futaki1983} and then generalized in \cite{Ding1992} and \cite{donaldson2002} as an obstruction to constructing metrics: its vanishing is a necessary condition of the existence of K\"ahler-Einstein metrics on Fano varieties. 
For general compact complex manifolds, it is conjectured that K-stability is equivalent to the existence of constant scalar curvature K\"ahler (cscK) metric.

In \cite{Collins:2012dh,2015arXiv151207213C}, the notion of K-stability was extended to any Sasakian manifold, including irregular ones. It was shown that if a Sasakian manifold $S$ with Reeb vector field $\zeta$ has a constant scalar curvature metric, then its cone $(\text{Cone}(S),\zeta)$ is K-semistable (see Definition \ref{def:k-semi}).
In particular, we can use Hilbert series (HS) to compute Futaki invariants. 
For an affine variety $X\subset\mathbb{C}^n$ cut out by some $I\subset\mathbb{C}[x_1,\dots,x_N]$ such that $X=\text{Spec}(R)$, where $R=\mathbb{C}[x_1,\dots,x_N]/I$, the symmetry/Reeb field $\zeta\in\mathfrak{t}$ acts on the functions on $X$ with positive weights, where $\mathfrak{t}$ is the Lie algebra of the torus action $T\subset\text{Aut}(X)$. Then we can write the HS with respect to the weighting of $\zeta$ (strictly, we should think of the HS as being associated to the weighted projective variety obtained from the projectivization of the affine variety, keeping the weights as multi-degrees).
To see if there exists a destabilizing ring which has a larger symmetry, we perturb the HS with a test symmetry $\eta$ by considering $(\zeta+\epsilon\eta)$. The information of the grading induced by $\eta$ is reflected by the coefficients (and derivatives thereof) in the Laurent expansion for the perturbed HS. With this data, we may follow the standard algebro-geometric set-up to compute the Futaki invariant.

Such idea can then be applied to various aspects in physics. It was shown that the Lichnerowicz obstruction in \cite{Gauntlett:2006vf} is in fact the problem of K-semistability for deformations arising from Rees algebras of principal ideals. Moreover, K-(semi)stability for product test configurations is equivalent to volume minimization. 
In light of AdS/CFT, this is then related to $a$-maximization \cite{Martelli:2005tp}. For a general test configuration induced by $\eta$, if we find some destabilizing ring at the central fibre (i.e., the flat limit of the test configuration) whose symmetry is $\zeta(\epsilon)$ parameterized by $\epsilon$, then following \cite{Collins:2016icw}, the Futaki invariant is equal to the derivative of $a_0(\zeta(\epsilon))$ with respect to $\epsilon$, where $a_0(\zeta(\epsilon))$ is the leading coefficient in the Laurent expansion for the HS of the destabilizing ring weighted by $\zeta(\epsilon)$. It turns out that this $a_0(\zeta(\epsilon))$ is inversely proportional to the central charge of the destabilizing chiral ring. Hence, K-stability, serving as some generalized $a$-maximization, is naturally related to the conformality of supersymmetric gauge theories.

The paper is organized as follows. In \S\ref{chiralgauge}, we first give a brief review on chiral rings and K-stability. Then we will present a quick formula to compute the Futaki invariant, noticing that there could be problems in the computations following the usual steps. We will show how to resolve these problems by modifying the numerators in the HS, and also try to simplify the process of handling test symmetries by some rescaling. In \S\ref{examples}, we will illustrate the ideas and computations with various 4d $\mathcal{N}=1$ examples, attempting to extend the calculations beyond hypersurface singularities and theories of D-branes probing Calabi-Yau (CY) manifolds in \cite{Collins:2016icw}. In Appendix \ref{gb}, we will review Gr\"obner basis which is a useful tool in obtaining the HS and hence in our calculations.

\section{Chiral Rings of Supersymmetric Gauge Theories}\label{chiralgauge}
We shall focus on the chiral rings of (3+1)-dimensional SCFT \cite{Cachazo:2001sg,Cachazo:2002ry,Cachazo:2003yc} for whose supersymmetry we will write in $\mathcal{N}=1$ language. 
In short, this is simply the set of operators $\mathcal{O}_i$ which are ``holomorphic'' in that they are annihilated by the supercharges $\bar{Q}_{\dot{\alpha}}$ so that they are defined modulo the cohomolgy thereof; hence there exists an operator $\chi$ such that
\begin{equation}
\mathcal{O}_i\sim\mathcal{O}_i+\left[\bar{Q}_{\dot{\alpha}},\chi\right] \ .
\end{equation}
The ring structure follows from the fact that (1) there is an identity operator $\mathcal{O} = \mathbb{I}$, (2) the sum and product of two chiral operators remain chiral, and (3) the structure constant is that for the (spacetime independent) OPE for the VEVs: $\mathcal{O}_i\mathcal{O}_j=\sum\limits_kC^k_{ij}\mathcal{O}_k$. In fact, this ring is a (finite) commutative ring with identity.

Computationally, the \emph{classical} chiral ring can be determined as follows. We have a superpotential $W$, which is a holomorphic polynomial in $\mathcal{O}_i$, each of which can be thought of as a matrix operator in an appropriate representation of the gauge group, with over-all trace. 
Consider all (complex) components $\phi_i$ of all the $\mathcal{O}_i$, and work over the polynomial ring $R=\mathbb{C}[\phi_i]$. The F-terms, constituted by the partial derivatives of $W$ with respective to $\phi_i$, can be thought of as the Jacobian ideal $J=\left<\partial_{\phi_i}W\right>\subset R$. The chiral ring can then be thought of as the quotient ring $R/J$ (giving us the ``master space'' \cite{Forcella:2008bb}), and then quotiented further by any polynomial relations which arise from the traces, such as those obeyed by Newton relations. 
For example, for $SU(N)$ theory with a chiral field $\Phi$ in the adjoint, the chiral ring is freely generated by the single-trace operators $\text{tr}(\Phi^i)$ for $i=0,1,2,\dots,N-1$ because any $\text{tr}\left(\Phi^{i>N}\right)$ can be written as Newton polynomials of the former and any multi-trace operator is just products of these single-traces.

The above should be compared and contrasted with the calculation of the \emph{classical} vacuum moduli space (VMS), which is the GIT quotient of $J$ by the complexified gauge group \cite{Luty:1995sd}.
Computationally, this is done by considering the minimal set of gauge invariant operators (GIOs) $G_j$ in the theory, each being a single-trace operator, and thus a polynomial in the $\phi_i$. Then the classical VMS is the image of quotient ring $R/J$ under the map $\{D_j\}$ into $S = \mathbb{C}[D_j]$ \cite{Gray:2006jb,He:2014oha,Hauenstein:2012xs}. Importantly, in AdS/CFT, this VMS is nothing more than the Calabi-Yau variety $X$ which a single brane probes and whose world-volume gauge theory is the SCFT; for $N$ parallel stack of D-branes, the VMS is the $N^{\text{th}}$ symmetric product of $X$. 

It should be emphasized that the classical chiral ring and the VMS both receive quantum corrections due to strongly coupled effects such as instantons. Algebro-geometrically, the correction often corresponds to a complex structure deformation. For example, in $\mathcal{N}=1$ SQCD, the classical chiral operators are the mesons $M^i_j = Q^i_a\tilde{Q}^a_j$ and baryons $B^{i_1\dots i_N}=\epsilon_{a_1\dots a_N}Q_{i_1}^{a_1}\dots Q_{i_N}^{a_N}$, $\tilde{B}=\epsilon_{a_1\dots a_N}\tilde{Q}_{i_1}^{a_1}\dots\tilde{Q}_{i_N}^{a_N}$ in terms of the quarks $Q_i$ and $\tilde{Q}_i$, with the famous relation for the VMS: $B^{i_1\dots i_N}\tilde{B}_{j_1\dots j_N}=M^{[i_1}_{j_1}\dots M^{i_N]}_{j_N}$. Interestingly, in \cite{Gray:2008yu}, it was shown that all the classical VMSs are affine Calabi-Yau (Gorenstein) singularities.

\subsection{R-Charges and $a$-Maximization}\label{amax}
The SCFT of our interest is in general the IR fixed point under renormalization group flow of some UV gauge theory. 
It is usually difficult to determine the exact U(1)$_R$ symmetry of an SCFT. In the spirit of Zamolodchikov's $a$-theorem for (1+1)-dimensional CFTs, the analogue $a$-theorem was beautifully developed by \cite{Intriligator:2003jj} for (3+1)-dimensional SCFTs.
The geometrical version of this in terms of $Z$-minimization of the Sasaki-Einstein horizon area in the dual AdS picture was given by \cite{Martelli:2005tp,Martelli:2006yb} which nicely applies to arbitrary quiver gauge theories for branes probing toric Calabi-Yau varieties.
The explicit method of computation for toric CYs was given in \cite{Butti:2005vn} and
an algorithmic phrasing thereof in the dimer/tiling language, in \cite{Hanany:2011bs}.

We summarize the methodology of finding the exact R-charges as follows:
\begin{itemize}
	\item To each operator (field) $\mathcal{O}_i$ assign a trial R-charge $R_i$ (this will be related to the conformal dimension as $\Delta_i=\frac32 R_i$);
	
	\item Define the \emph{conformal manifold}
	\begin{equation}
	\mathcal{M} = \left\{ R_i > 0 \quad : \quad 
	\sum\limits_i R_i = 2 \ , \ \ 
	\sum\limits_i (1 - R_i) = 2
	\right\}.
	\end{equation}
	The first sum is taken over the charges of the operators for \emph{each monomial term} in the superpotential $W$; this is simply to ensure that $W$ has homogeneously R-charge 2 so that it can be integrated in superspace against $\int\text{d}\theta^2$. The second sum means that if we have the gauge group which is a direct product over factors, such as in quiver theories, for \emph{each group factor}, we need to sum over the R-charges of all the fields under this group; note for adjoint fields, we need to sum over twice since they can be thought of as bi-directional arrows in the quiver. In other words, we consider all fundamentals and anti-fundamentals charged under the gauge group factor and adjoints are considered as both.
	
	\item Consider the trial $a$-function
	\begin{equation}
	a(R_i)=\frac{9}{32}\left(N_G+\sum\limits_i (R_i-1)^3\right),
	\end{equation}
	where now we sum over \emph{all} operators, and $N_G$ denotes the number of gauge groups, which comes from the contributions from the gauginos. Note that the $-1$ may look slightly unfamiliar, but the usual formula $a = \frac{3}{32} \left( 3( \text{tr} R^3) - \text{tr} R \right)$ has the trace over all the fermion representations, which is 1 less than the bosons in the same multiplet (cf. eq (1.9) of the original paper \cite{Intriligator:2003jj}).
	
	\item Maximize $a(R_i)$ on constraints imposed by the conformal manifold $\mathcal{M}$ and this will give the correct R-charges. There are general statements as to the uniqueness of this maximum \cite{Kato:2006vx}.
\end{itemize}

As an example, $\mathcal{N}=4$ SYM has three adjoint fields $X,Y,Z$ charged under the single gauge group U($N$), with superpotential $W=\text{tr}(XYZ-XZY)$. We thus have three R-charges $R_X, R_Y, R_Z$ and $\mathcal{M}$ is given by the constraints $R_X + R_Y + R_Z = 2$ and $2(1 - R_X) + 2(1 - R_Y) + 2(1 - R_Z) = 2$ with $R_X, R_Y, R_Z > 0$. Maximizing $a = \frac{9}{32}\left(1+(R_X-1)^3 + (R_Y-1)^3 + (R_Z-1)^3\right)$ on $\mathcal{M}$ gives the familiar $R_X=R_Y=R_Z=\frac23$.

\subsection{Hilbert Series}\label{hs}
One of the most important quantities which characterize an algebraic variety $X$ is the Hilbert series. 
The relevance of computing the HS in relation to the volume of the Sasaki-Einstein base in toric AdS/CFT has been the beautiful work of \cite{Martelli:2005tp,Martelli:2006yb,Bergman:2001qi}. In parallel, a plethystic programme was established \cite{Benvenuti:2006qr,Feng:2007ur} addressing the key problem of counting GIOs in gauge theory (q.v.~ \cite{Gray:2008yu,Forcella:2008bb,Hanany:2015via,Braun:2012qc,Rodriguez-Gomez:2013dpa,Cremonesi:2014kwa}). 
Moreover, its properties have also been exploited to study the phenomenology of the standard model, ranging from question of vacuum structure to operator selection \cite{Hanany:2010vu,Gray:2006jb,He:2014loa,He:2014oha,He:2015rzg,Henning:2015daa,Lehman:2015coa,Xiao:2019uhh}.

We recall that for a variety $X$ in $\mathbb{C}[x_1,...,x_k]$, the HS is the generating function for the dimension of the graded pieces:
\begin{equation}
\text{HS}(t;X)=\sum\limits_{i=0}^{\infty}\left(\dim_{\mathbb{C}}X_i\right)t^i,
\end{equation}
where $X_i$, the $i^{\text{th}}$ graded piece of $X$ can be thought of as the number of independent degree $i$ (Laurent) polynomials on the variety $X$. The most useful property of HS is that it is a rational function in $t$ and can be written in 2 ways:
\begin{equation}\label{hs12}
\text{HS}(t; X) = \left\{
\begin{array}{ll}
\frac{Q(t)}{(1-t)^k} \ , & \mbox{HS of first kind} \ ;\\
\frac{P(t)}{(1-t)^{\dim(X)}} \ , & \mbox{HS of second kind} \ . 
\end{array}
\right.
\end{equation}
Importantly, both $P(t)$ and $Q(t)$ are polynomials with \emph{integer} coefficients and the powers of the denominators are such that the order of the pole captures the dimension of the variety and the embedding space $\mathbb{C}^k$ within which $X$ is an algebraic variety, respectively for the first and second kind.

Let us summarize a few key properties of the HS which we will need:
\begin{itemize}
	\item It is {\em not} a topological invariant and does depend on embedding and choice of grading/weighting for the coordinate ring for $X$. The weight comes from a choice of a symmetry/Reeb vector field $\zeta$ of the theory. Typically, we choose the U(1)$_R$ symmetry of the SCFT to weight the fields, and, thence the GIO variables of $X$;
	
	\item Written in the second kind, $P(1)$ equals to the degree of the variety;
	
	\item Also in the second kind, if $P(t)$ is \emph{palindromic}, then Stanley's theorem says this is equivalent to $X$ being Gorenstein \cite{STANLEY197857}, which for our purposes can be taken to mean affine Calabi-Yau;
	
	\item
	A Laurent expansion for the Hilbert series of second kind in \eqref{hs12}
	can be developed, as a partial fraction expansion:
	\begin{equation}\label{laurent}
	\text{HS}(t; X)= \frac{V_n}{(1-t)^{n}} + \dots
	\frac{V_3}{(1-t)^{3}} + \frac{V_2}{(1-t)^{2}} + \frac{V_1}{1-t}
	+ V_0 + \mathcal{O}(1-t) \ ,
	\end{equation}
	where we see explicitly that the Hilbert series is a rational
	function and the degree of its most singular pole is the dimension of
	$X$.
	
	In the case of $X$ being a toric Calabi-Yau variety of dimension 3 (such as in the vast majority of known cases of AdS$_5$/CFT$_4$), the coefficients $V_{0,1,2,3}$ are related directly to the Reeb vector of $X$ so that $V_3$ is the volume of the
	spherical Sasaki-Einstein horizon\footnote{The relation to the Reeb vector, at least for toric $X$, is as follows \cite{Martelli:2006yb}. Refine the generating function into tri-variate (this can always be done for toric $X$), in terms of $t_{i=1,2,3}$ and set $t_i:=\exp(-b_iq)$ where $\vec{b} = (b_1, b_2, b_3)$ is the Reeb vector for the 3 isometries of $X$ as a toric variety. Then Laurent expand $f(t_1,t_2,t_3)$ near $q \to 0$ to compare with \eqref{laurent}.}.
	
	\item
	In the notation of \cite{Collins:2016icw}, suppose the underlying (Calabi-Yau) geometry (VMS) is $X$, of complex dimension $n=3$, we have a U(1)$_R$ symmetry $\zeta$ with the associated trial central charge $a(\zeta)$, we perform the Laurent expansion of the Hilbert series as
	\begin{equation}\label{sLaurent}
	\text{HS}(t = e^{-s}, \zeta;X) = \frac{a_0(\zeta)}{s^3} + \frac{a_1(\zeta)}{s^2} + \dots
	\end{equation}
	Then, we have that
	\begin{itemize}
		\item the coefficient $a_0$ is proportional to the (normalized) volume of the base over which $X$ is a cone (for example, $X = \mathbb{C}^{3} = \text{{Cone}}(S^5)$);
		\item the trial $a$-charge (of order $N^2$) is given by
		\begin{equation}
		a(\zeta) = \frac{27 N^2}{32} \frac{1}{a_0(\zeta)};
		\end{equation}
		\item the holomorphic volume $(3,0)$-form $\Omega$ (from the Calabi-Yau condition of $X$) will be chosen to have charge 2, which implies that $a_0 = a_1$;
		\item the coefficient $a_0(\zeta)$ is a convex function in the symmetry generators \cite{Martelli:2005tp}.
	\end{itemize}

\item For \emph{complete intersection} varieties, i.e., the codimension of $X$ being exactly equal to the number of defining polynomials, the HS is relatively easy to construct \cite{Benvenuti:2006qr,Feng:2007ur}. In particular, the simplest case of a complete intersection is that of a single defining equation and $X$ being codimension 1, viz., a hypersurface. For example, consider the quadric hypersurface $Q = \{x^2 + y^2 + z^2 + w^2 = 0\}$ in $\mathbb{C}^4$, otherwise known as the conifold singularity as a local Calabi-Yau threefold. Suppose we weigh the variables as $W(x,y,z,w) = (1,1,1,1)$, then we have 4 generators (variables), each of degree 1, obeying the one quadratic defining relation, of degree 2. For each generator we place a factor of $(1 - t^W)$ in the denominator, and for each relation of degree $d$, we place a factor of $(1 - t^d)$ in the numerator.
Therefore, the HS here is simply $\text{HS}(t;Q)=\frac{1-t^2}{(1-t)^4}$.

In fact, one can define a pair of inverse functions \cite{Feng:2007ur}, the \emph{plethystic exponential} PE$[f(t)]$ and the \emph{plethystic logarithm} PL$[f(t)]$ for any analytic function $f(t)$ affording Taylor series about 0:
\begin{equation}
f(t) = \sum\limits_{n=0}^\infty a_n t^n \quad \Rightarrow
\left\{
\begin{array}{l}
\text{PE}[f(t)] = \exp\left( \sum\limits_{n=1}^\infty \frac{f(t^n) -
	f(0)}{n} \right) =
\prod\limits_{n=1}^\infty (1-t^n)^{- a_n}
\\
\text{PL}[f(t)] = \sum\limits_{k=1}^\infty \frac{\mu(k)}{k} \log (f(t^k))
\end{array}
\right.
\end{equation}
where $\mu(k)$ is the M\"obius function, which for an integer $k$ is equal to 0 if $k$ has repeated prime factors, equal to 1 if $k = 1$ and equal to $(-1)^n$ if $k$ is a product of $n$ distinct primes. That the above pair are indeed inverses of each other is non-trivial and involves the arithmetic properties of $\mu$.

The remarkable fact is that (though it has poles at $t=1$) the HS is analytic about $t=0$ and can be used as the functional argument of PE and PL. Indeed, HS$(t;X)$ for $X$ being the supersymmetric vacuum moduli space of the SCFT is the generating function for the single-trace operators in the chiral ring and PE$[\text{HS}(t;X)]$ counts the multi-trace operators. Moreover, PL$[\text{HS}(t;X)]$ is a polynomial for complete intersections and explicitly counts the generators (the first positive terms) and relations (the first negative terms) for $X$ of each degree\footnote{For non-complete intersections, there are terms of higher orders known as \emph{syzygies} that enumerates relations among basic relations and generators.}. For our above conifold example, $\text{PL}[\text{HS}(t;Q)]=4t-t^2$, signifying 4 degree-one generators obeying 1 quadratic relation.

\item It should be emphasized that the generic variety, and chiral ring, is \emph{not} complete intersection and the presentation of the generators and relations could be rather complicated.
In such situations, the most standard method is to compute the \emph{Gr\"obner basis} of $X$. The advantage of the Gr\"obner basis method is that it is algebraic and algorithmic. We describe this in more detail in Appendix \ref{gb}. On the other hand, as we are considering the Higgs branch, we can also use another method, namely the \emph{Molien-Weyl integral}, to compute HS. For a detailed treatment, readers are referred to \cite{Benvenuti:2006qr}.
\end{itemize}

\subsection{Flat Limits and Central Fibres}\label{flat}
As mentioned in \S\ref{intro}, (K-)stability and the Futaki invariant are related to the existence of a destabilizing ring for $X$. We start with some \emph{test configuration} $X_t$, that is, $X$ with a one-parameter subgroup $\eta(t):\mathbb{C}^*\hookrightarrow\text{GL}(m,\mathbb{C})$. For any polynomial $f$, in our convention, we have $(\eta(t)\cdot f)(x_0,\dots,x_m)=f(\eta(t)x_0,\dots,\eta(t)x_m)$. We will always assume that $\eta(t)$ is \emph{diagonal} under a unitary change of basis. The test configuration now has the ring $\mathbb{C}[x_0,\dots,x_m]/I_t$ with $I_t=\{\eta(t)\cdot f|f\in I\}$, where $I$ is the ideal defining the ring of $X$. Then to get the \emph{central fibre}, we need to take the \emph{flat limit} defined as follows (see Appendix \ref{gb} for details on initial ideals and polynomial ordering).
\begin{definition}
	For any $f\in I$, we find the initial polynomial $\textup{in}(f)$ with respect to the ordering defined by $\eta(t)$ such that $\textup{in}(f)$ is the lowest weight polynomial. Then the flat limit of $I_t$ is $I_0=\lim\limits_{t\rightarrow0}I_t=\{\textup{in}(f)|f\in I\}$.
\end{definition}
Notice that, however, following \cite{2015arXiv151207213C,Collins:2016icw}, it should be a \emph{partial} ordering rather than a total ordering. For instance, consider the conifold $w^2+x^2+y^2+z^2=0$. If we have $\eta(t)\cdot(w,x,y,z)=(tw,x,y,z)$, then the test configuration is $t^2w^2+x^2+y^2+z^2=0$. Taking the flat limit gives the central fibre $x^2+y^2+z^2=0$, rather than a single monomial. On the other hand, if we consider $\eta(t)\cdot(w,x,y,z)=(t^{-1}w,x,y,z)$, i.e., the test symmetry $\eta$ with charges $(-1,0,0,0)$, we would get $w^2=0$.

It is also worthing noting that for more general cases, if we simply take the initial polynomials of the generators of the ideal, we may get a smaller ideal than the flat limit \cite{szekelyhidi2014introduction}\footnote{The reason behind it is actually related to the syzygies. For more details, see \cite{1993alg.geom..4003B,artin1976lectures}.}. To get the exact flat limit, the strategy is to compute the Gr\"obner basis. Let us consider the twisted cubic curve example in \cite{1993alg.geom..4003B}, where $I=\langle f_1,f_2,f_3\rangle$ for $f_1=w^2-xy,~f_2=wy-xz,~f_3=wz-y^2$, and the action is $\eta(t)=(t^{-16},t^{-4},t^{-1},1)$. The test configuration is
\begin{equation}
	\eta f_1=t^{-32}w^2-t^{-5}xy,~\eta f_2=t^{-17}wy-t^{-4}xz,~\eta f_3=t^{-16}wz-t^{-2}y^2.
\end{equation}
Naively, the flat limit is generated by $w^2,~wy,~wz$. However, if we consider the Gr\"obner basis for $f_i$, we have
\begin{equation}
	w^2-xy,~wy-xz,~wz-y^2,~xz^2-y^3.
\end{equation}
Hence, the flat limit should really be generated by $w^2,~wy,~wz,~xz^2$.

\subsection{Futaki Invariant and K-Stability}\label{futaki}
Let us start with the (polarized) ring $(X,\zeta)$ with symmetry $\zeta$. Throughout, by ``polarized'' we mean that the ring is also equipped with a Reeb symmetry.
Also note, by slight abuse of notation, that we will use $X$ for varieties and associated coordinate rings interchangeably.
Then to find out whether there would be a ring destabilizing $X$, we need to consider some test symmetry $\eta$.
As aforementioned, this is done by considering some test configuration $X_t=\mathbb{C}[x_i]/I_t$ induced by the test symmetry, and then taking the flat limit $t\rightarrow0$ to get the central fibre $X_0=\mathbb{C}[x_i]/I_0$. For general $t$, $X_t$ would be isometric to $X$ while $X_0$ may or may not be trivial.

From \cite{szekelyhidi2014introduction}, we know that the total weight $w_k$ of the action on the (sufficiently high) degree $k$ piece of our graded ring can be written as a polynomial
\begin{equation}
	w_k=b_0k^{n}+b_1k^{n-1}+\dots \ ,
\end{equation}
where from \cite{Collins:2012dh}, we learn that (up to a positive constant dependent only on the dimension $n$)\footnote{In fact, up to some convention, the $a_i$'s also act as leading and subleading coefficients of a polynomial, namely the dimension $d_k$ of the degree $k$ piece of the graded ring: $d_k=a_0k^n+a_1k^{n-1}+\dots$, which is nothing but the Hilbert function of $X$.}
\begin{equation}
	b_i=-\frac{1}{n-i}\text{D}_\epsilon a_i(\zeta+\epsilon\eta)\bigg|_{\epsilon=0} \ .
	\label{bi}
\end{equation}
The Futaki invariant is then defined as\footnote{There is also a differential geometric definition of Futaki invariant. Specifically, for a smooth $n$-dimensional normal variety $X$ (the generalizations allow $X$ to be singular) with K\"ahler form $\omega\in[c_1(T_{X})]$ and Ricci potential $h_\omega$ so that $\text{Ric}(\omega)-\omega=\frac{i}{2\pi}\partial\bar{\partial}h_\omega$ where Ric$(\omega)$ is the Ricci form. Then the Futaki invariant, for some holomorphic vetor field $v$ on $X$, is $F_{c_1(T_{X})}(v)=\int_{X}v(h_\omega)\omega^n$. Since it is a character on the Lie algebra of $v$ and independent of the choice of $\omega$, this is an holomorphic invariant \cite{Linotes}. One can show that if $X$ is smooth and the $\mathbb{C}^*$-action is induced by a holomorphic vector field, then \eqref{futdef1} is the sames as the differential geometric Futaki invariant \cite{szekelyhidi2014introduction}.} \cite{szekelyhidi2014introduction}
\begin{equation}
	F(X;\zeta,\eta)=\frac{a_1}{a_0}b_0-b_1 \ .
	\label{futdef1}
\end{equation}
There is also an equivalent definition in \cite{Collins:2012dh,2015arXiv151207213C}\footnote{Notice that due to different conventions of $a_0$ and $a_1$, our definition here should agree with the definition in \cite{Collins:2012dh,2015arXiv151207213C,szekelyhidi2014introduction} up to some positive constant depending only on dimension.}:
\begin{equation}
	F(X;\zeta,\eta)=\text{D}_\epsilon a_0(\zeta+\epsilon\eta)+na_0\text{D}_\epsilon\frac{a_1(\zeta+\epsilon\eta)}{a_0(\zeta+\epsilon\eta)}\bigg|_{\epsilon=0}\ ,
	\label{futdef2}
\end{equation}
where D$_\epsilon$ is defined in \eqref{Futaki1} below.
We remark that the Futaki invariant in its original context, was in terms of a integral as detailed in the footnote, due to the purely algebraic recasting above, it is sometimes referred to as the Futaki-Donaldson invariant.

Algorithmically, our Futaki invariant can be determined as follows \cite{Collins:2016icw}:
\begin{itemize}
	\item For a symmetry/weighting $\zeta$ of the variables of $X$ such that the holomorphic top form has charge/weight 2, compute the HS (thus in particular $a_0(\zeta) = a_1(\zeta)$ in our convention);
	\item Find a test symmetry $\eta$ of $X$, expressed as a vector of weights\footnote{Technically, $\eta$ is a square matrix, but as we will see, it is always assumed to be diagnolizable.}, as $\zeta$;
	\item Consider the possible U(1)$_R$ symmetry, for some small $\epsilon>0$ (so that the central fibre from the test symmetry $\epsilon(\eta-a\zeta)$ is the same as the one from $\eta$),
	\begin{equation}
	\zeta(\epsilon)=\zeta+\epsilon(\eta-a\zeta)=(1-a\epsilon)\zeta+\epsilon\eta,
	\end{equation}
	where $a$ can be obtained from
	\begin{equation}
		a=\frac{1}{a_0(\zeta)}\left(\frac{\text{d}a_1(\zeta+\epsilon\eta)}{\text{d}\epsilon}-\frac{\text{d}a_0(\zeta+\epsilon\eta)}{\text{d}\epsilon}\right)\bigg|_{\epsilon=0}.
	\end{equation}
	\item With respect to this new weighting, compute the HS and perform the usual Laurent expansion \eqref{sLaurent} to extract the coefficients $a_0\left( \zeta(\epsilon) \right) = a_1\left( \zeta(\epsilon) \right)$;
	\item The Futaki invariant is obtained by
	\begin{equation}
	F(X;\zeta,\eta) = \frac{\partial}{\partial \epsilon}a_0\left( \zeta(\epsilon) \right)\bigg|_{\epsilon = 0}=:\text{D}_\epsilon a_0(\zeta(\epsilon))|_{\epsilon=0}.\label{Futaki1}
	\end{equation}
\end{itemize}

As argued in \cite{Collins:2016icw}, \eqref{Futaki1} is equivalent to the original definition of Futaki invariant in \cite{donaldson2002} by considering
\begin{eqnarray}
	F&=&\text{D}_\epsilon a_0\bigg(\zeta(\epsilon)=\zeta+\epsilon(\eta-a\zeta)\bigg)\bigg|_{\epsilon=0}\nonumber\\
	&=&(\eta-a\zeta)\cdot a_0'|_{\epsilon=0}\nonumber\\
	&=&\eta\cdot a_0'-a\zeta\cdot a_0'|_{\epsilon=0}\nonumber\\
	&=&\text{D}_\epsilon a_0(\zeta+\epsilon\eta)+\frac{1}{a_0}\left(\frac{\text{d}a_1(\zeta+\epsilon\eta)}{\text{d}\epsilon}-\frac{\text{d}a_0(\zeta+\epsilon\eta)}{\text{d}\epsilon}\right)na_0\bigg|_{\epsilon=0}\nonumber\\
	&=&\text{D}_\epsilon a_0(\zeta+\epsilon\eta)+na_0\text{D}_\epsilon\frac{a_1(\zeta+\epsilon\eta)}{a_0(\zeta+\epsilon\eta)}\bigg|_{\epsilon=0},\label{Futaki2}
\end{eqnarray}
where we have used $\zeta\cdot a_i'=\text{D}_\epsilon a_i(\zeta+\epsilon\zeta)=-(n-i)a_i(\zeta)$ to get the fourth line, and the last equality is the quotient rule of derivatives with $\eta\cdot a_0|_{\epsilon=0}=\eta\cdot a_1|_{\epsilon=0}=a_0$. As we can see, the result obtained in \eqref{Futaki2} is exactly \eqref{futdef2}.

Following the third line in \eqref{Futaki2}, it is straightforward that $F$ is linear with respect to the test symmetry. For the first term, we have $(s\eta_1+\eta_2)\cdot a_0'=s\eta_1\cdot a_0'+\eta_2\cdot a_0'$ ($s>0$). Hence, it is equivalent to showing that $a$ is linear with respect to the test symmetry, which is then equivalent to showing that $\text{D}_\epsilon a_i(\zeta+\epsilon\eta)$ is linear. This is certainly true as $\text{D}_\epsilon a_i(\zeta+\epsilon(s\eta_1+\eta_2))=(s\eta_1+\eta_2)\cdot a_i'=s\eta_1\cdot a_i'+\eta_2\cdot a_i'$.

Moreover, from the fourth line in \eqref{Futaki2}, we also have
\begin{equation}
	F=n\text{D}_\epsilon a_1(\zeta+\epsilon\eta)-(n-1)\text{D}_\epsilon a_0(\zeta+\epsilon\eta)|_{\epsilon=0}.\label{Futaki3}
\end{equation}
Inserting \eqref{bi}, we find that this is the same as definition \eqref{futdef1} (up to some positive coefficient). Therefore, \eqref{Futaki1}$\sim$\eqref{Futaki3} all give the same answer and we can use them interchangeably.

As K-stability depends on the sign of Futaki invariant, we can almost introduce its definition. However, whether a test configuration is trivial still needs to be determined especially when $F=0$. A test configuration was initially defined to be trivial when the central fibre is biholomorphic to $X$. However, as shown in \cite{2011arXiv1111.5398L}, there exist non-trivial test configurations (which are trivial in codimension 1) satisfying biholomorphicity. To avoid such pathological cases, one has to restrict to normal (or $S_2$) test configurations when $X$ is normal (or $S_2$). Here, following \cite{szekelyhidi2014introduction}, we will use an alternative way to determine the K-stability when $F$ vanishes without the normality condition. In particular, one can introduce the norm $||\eta||$ by considering the infinitesimal generator $A_k$ of the $\mathbb{C}^*$-action on the degree $k$ piece of the ring. It is not hard to see that $\text{tr}(A_k)=w_k$. We can also define $c_0$, which is also a constant with respect to degree $k$, by 
\begin{equation}
	\text{tr}(A_k^2)=c_0k^{n+1}+\dots,
\end{equation}
and it is shown in \cite{Collins:2012dh} that (up to a positive constant same as in $b_0$)
\begin{equation}
	c_0=\frac{1}{n(n+1)}\text{D}_\epsilon^2a_0(\zeta+\epsilon\eta)|_{\epsilon=0}.
\end{equation}
Then we can define the norm as
\begin{equation}
	||\eta||^2=\left\{
	\begin{array}{ll}
	0,&I_0\cong I_{t\neq0};
	\\
	c_0-\frac{b_0^2}{a_0},~&\text{otherwise}.
	\end{array}
	\right.\label{norm}
\end{equation}

Thus defined, the notion of K-stability is clear:
\begin{definition}\label{def:k-semi}
	The ring $(X,\zeta)$ is K-semistable if for any test symmetry $\eta$, we have $F(X;\zeta,\eta)\geq0$. If in addition $F=0$ only when the norm vanishes, then the ring is K-stable.
\end{definition}
Let us have a closer look at the case with $F=0$. A trivial test configuration (which leads to $F=0$) for a K-stable ring should always have a vanishing norm. In the usual K-stability context, a well-defined triviality should be the equivalent to the norm being zero. However, as we will see below, besides the second line in \eqref{norm}, the first line is also necessary since there could be trivial configurations with non-zero values for the second line\footnote{In fact, there are various conventions to define K-stability in various literature. In some texts dealing with Fano manifolds, the ``K-stability'' we are considering here would be called ``K-polystability'' which could be subtlely different. Here, we will adopt the convention so that the trivial test configurations arise from automorphisms will automatimatically have norm zero. We would like to thank G\'abor Sz\'ekelyhidi for helpful advice on this.}.

It is then the conjecture of \cite{Collins:2016icw} saying that
\begin{conjecture}
	  The ring $(X,\zeta)$ is the chiral ring of an SCFT iff $X$ is K-stable.\label{conj1}
\end{conjecture}
As we will see, there seems to exist a counterexample where this K-stability criterion would not work. However, this is still possible to be true for a sub-class of supersymmetric theories such as the worldvolume theories of D3-branes probing CY$_3$.

\subsection{Futaki Invariants for Non-Complete Intersections}\label{noncomplete}
For complete intersections, the denominators of the HS encode the charges of the coordinates/generators. With the aforementioned method, the Futaki invariants can then be quickly computed as in \cite{Collins:2016icw} since we can directly add the test charges to the corresponding terms in the denominator of HS. Here, we propose a method allowing us to obtain the Futaki invariants with Hilbert series which also works for general varieties.

We would like to know which factor in the HS our test symmetry can act on, but for non-complete intersections this piece of information is hidden (especially when we derive the HS from quivers in physics). The denominator simply encodes the dimension of the variety while the numerator contains other complicated data. Therefore, we can naturally use the plethystic logarithm to reveal the information we need.

We start with a general HS and take its PL whose first positive terms tell us all the generators at different degrees. For instance, if we have a generator of order $k$ (and hence with weight/charge $k$), then we multiply the HS with $(1-t^k)$ on its denominator and numerator:
\begin{equation}
	\text{HS}_{\zeta}=\frac{1-t^k}{1-t^k}\text{HS}=\frac{1-t^k}{1-t^k}\times\frac{P(t)}{\left(1-t^m\right)^{\text{dim}(X)}}.
\end{equation}
As we write out the specific generator explicitly in the denominator, as in the complete intersection case, we can easily get the HS for test symmetry $\eta$ where only the generator at order $k$ has non-vanishing charge:
\begin{equation}
	\text{HS}_{\zeta+\epsilon\eta}=\frac{1}{1-t^{k+\epsilon\eta}}\times\frac{\left(1-t^k\right)P(t)}{\left(1-t^m\right)^{\text{dim}(X)}}.
\end{equation}
Now we can immediately get $a_0(\zeta+\eta\epsilon)$ and $a_1(\zeta+\eta\epsilon)$ as usual. Then the Futaki invariant directly follows from \eqref{Futaki1}$\sim$\eqref{Futaki3}. If we use \eqref{Futaki1}, the Hilbert series for $\zeta(\epsilon)$ reads
\begin{equation}
	\text{HS}_{\zeta(\epsilon)}=\frac{\left(1-t^{k(1-a\epsilon)}\right)P\left(t^{(1-a\epsilon)}\right)}{\left(1-t^{k(1-a\epsilon)+\epsilon\eta}\right)\left(1-t^{m(1-a\epsilon)}\right)^{\text{dim}(X)}}.
\end{equation}
One may also check that for complete intersections, this approach reduces to the usual method before. We will see an example validating this approach on complete intersections in \S\ref{dP}.

To determine the stability, usually we need to consider quite a few test symmmetries. By the linearity discussed in \S\ref{futaki}, it suffices to compute the test symmetries $\eta_i$ with charge $\delta_{ij}$ for the $j^\text{th}$ generator. Any test symmetry and hence $F$ can be written as a linear combination of $\eta_i$'s (though crucially it still requires some work to figure out what kinds of linear combinations we want). In fact, we can use this to get Futaki invariants in a quicker way as follows.

Suppose we have a generator of order/charge $k$ under $\zeta$. Let us show that for the test symmetry with charge $(0,\dots,0,1,0,\dots,0)$, where only this generator of order $k$ has a non-vanishing charge, the Futaki invariant would have a simple expression. As usual, the HS has coefficient $a_i$ for the $s^{-(n-i)}$ term under expansion around $s=0$. Then with the test symmetry, we have
\begin{eqnarray}
	\text{HS}_{\zeta+\epsilon\eta}&=&\frac{\text{HS}_\zeta\times\left(1-\text{e}^{-ks}\right)}{1-\text{e}^{-(k+\epsilon)s}}\nonumber\\
	&=&\frac{a_0k}{(k+\epsilon)s^n}+\frac{k(\epsilon a_0+2a_1)}{2(k+\epsilon)s^{n-1}}+\dots
\end{eqnarray}
Since $a_0=a_1$, we have
\begin{equation}
	a_0(\zeta+\epsilon\eta)=\frac{a_0k}{k+\epsilon},~a_1(\zeta+\epsilon\eta)=\frac{a_0k(\epsilon+2)}{2(k+\epsilon)}.
\end{equation}
Now using (the second line in) \eqref{Futaki3}, we get
\begin{equation}
	F=n\frac{\text{d}}{\text{d}\epsilon}\frac{a_0k(\epsilon+2)}{2(k+\epsilon)}-(n-1)\frac{\text{d}}{\text{d}\epsilon}\frac{a_0k}{k+\epsilon}\bigg|_{\epsilon=0}=\frac{nk-2}{2k}a_0.
\end{equation}
Likewise, using \eqref{norm},
\begin{equation}
	||\eta||^2=\frac{(n-1)a_0}{n^2(n+1)k^2}.
\end{equation}
Incidentally, we can find that
\begin{equation}
	a=\frac{1}{a_0}\left(\frac{\text{d}}{\text{d}\epsilon}\frac{a_0k(\epsilon+2)}{2(k+\epsilon)}-\frac{\text{d}}{\text{d}\epsilon}\frac{a_0k}{k+\epsilon}\right)\bigg|_{\epsilon=0}=\frac{1}{2}.
\end{equation}
We can also write a general expression for general test symmetries. Suppose we have a test symmetry $\eta$ with charge $v_i$ for the $i^\text{th}$ generator which has order $k_i$, then
\begin{equation}
	a_0(\zeta+\epsilon\eta)=a_0\prod_i\frac{k_i}{k_i+v_i\epsilon},~a_1(\zeta+\epsilon\eta)=a_0\prod_i\frac{k_i(v_i\epsilon+2)}{2(k_i+v_i\epsilon)},
\end{equation}
and
\begin{equation}
	a=\frac{1}{a_0}\times\frac{a_0}{2}\sum_iv_i=\frac{1}{2}\sum_iv_i.
\end{equation}
The Futaki invariant is
\begin{equation}
	F=\sum_iv_i\frac{nk_i-2}{2k_i}a_0,\label{Futaki4}
\end{equation}
and the norm is
\begin{equation}
	||\eta||^2=\left\{
	\begin{array}{ll}
	0,&I_0\cong I_{t\neq0};
	\\
	\frac{(n-1)a_0}{n^2(n+1)}\left(\sum\limits_i\frac{v_i^2}{k_i^2}-\frac{2}{n-1}\sum\limits_{j<l}\frac{v_jv_l}{k_jk_l}\right),~&\text{otherwise}.
	\end{array}
	\right.\label{norm1}
\end{equation}

As an example, consider the orbifold $\mathbb{C}^3/\mathbb{Z}_5$ (1,2,2) studied in \cite{Benvenuti:2006qr,Bao:2020kji} with
\begin{equation}
\text{HS}=\frac{1-t^{2/3}+3t^{2}-t^{8/3}+3t^{10/3}-t^{14/3}+t^{16/3}}{\left(1-t^{2/3}\right)^3\left(1+t^{2/3}+t^{4/3}+t^2+t^{8/3}\right)^2}.
\end{equation}
Under Laurent expansion around $s=0$, we have $a_0=a_1=27/40$. 
Notice that here the {\em fractional powers} in the HS is just a consequence of our convention $a_0=a_1$. Hence, they do not have to equal the corresponding R-charges numerically.

The PL of HS reads
\begin{equation}
\text{PL(HS)}=3t^2+2t^{8/3}+7t^{10/3}-t^4-\dots,
\end{equation}
where we see that there are 3 generators of order 2, 2 generators of order $8/3$ and 7 generators of order $10/3$. Therefore, we can quickly get a general expression for Futaki invariant using \eqref{Futaki4}:
\begin{equation}
F=\frac{27}{40}(v_1+v_2+v_3)+\frac{243}{320}(v_4+v_5)+\frac{81}{100}(v_6+\dots+v_{12}),
\end{equation}
for test symmetry $\eta$ with charges $(v_1,v_2,\dots,v_{12})$. However, notice that this example is just for a pure calculation purpose: the orbifold here is actually a toric variety. As briefly aforementioned, for any toric singularity, there is no non-trivial test configuration because the number of $\mathbb{C}^*$-actions is already maximal \cite{Collins:2016icw,Fazzi:2019gvt}, or in other words, it has complexity zero. As a result, we should always expect the rings to be stable. We can also think of the quiver gauge theories which stay in the toric phase. Hence, there is no fractional brane that would prevent our theory from being conformal. On the other hand, for non-toric cases, we still need to find appropriate test symmetries to determine the stability.

\subsection{Test Symmetries}\label{testsym}
In practice, there could be a lot of possible test symmetries for us to consider. To guarantee stability, we need to exhaustively check all these Futaki invariants, which can be difficult. 
However, we could try to reduce the number of test symmetries we need to check. As argued in \cite{Fazzi:2019gvt}, for hypersurface singularities, especially for those with complexity one (i.e. having isometry U(1)$^{n-1}$) whose degeneration is toric, we can consider $X$ as a fibration over some Riemann surface, with the torus action acting on the fibre. Then the integer slopes of some piecewise-linear functions would help us find the correct test symmetries we want. See \cite{Fazzi:2019gvt,2015arXiv150704442I} for more details. In general, from the perspective of field theory by viewing $X_t$ as a deformation of $X_0$, it is also conjectured in \cite{Xie:2019qmw} that it should suffice to only consider the test configurations that remove one of the monomials for (isolated) hypersurface singularities.

For non-hypersurface singularities or even non-complete intersections, the above methods are not applicable (except that the toric varieties still have no non-trivial test configurations). First of all, we need to get the relations on which we can act with the one-parameter $\mathbb{C}^*$-subgroup and take the flat limit. This can again be found by taking the PL of HS, where the relations are given by the first negative terms, but we need the refined HS to get the exact relations. For instance, if we have \cite{Benvenuti:2006qr}
\begin{equation}
	\text{PL}\left(\frac{xy(1/q^2-1)}{(1-qx)(1-qy)(1-x/q)(1-y/q)}\right)=\frac{q}{x}+qx+\frac{q}{y}+qy-q^2,
\end{equation}
where $x,y,q$ are the fugacities. The defining equation is then given by $(q/x)(qx)=(q/y)(qy)=q^2$, viz, $uv=wz$, which is exactly the conifold.

As detailed in \S\ref{flat}, we should take the Gr\"obner basis of the relations to avoid generating a set smaller than the flat limit. Now when taking a test configuration, we always have some action $\eta(t)$ acting on these equations\footnote{Notice that for hypersurfaces, there is no need to find the Gr\"obner basis, and the coefficients in front of the terms in the equations do not matter.}. Then we will only keep the term(s) with lowest weight in each equation under the flat limit. In principle, there could be infinitely many $\eta$'s. However, there might be fewer cases due to the symmetries of the variables in the equation(s).

Moreover, as checking stability is equivalent to checking the positivity of Futaki invariants, and the sign of \eqref{Futaki4} is determined by $v_i$'s, the $v_i$-space would be divided into different areas which correspond to positive or negative Futaki invariants (recall that if $F=0$, we can check the norm). In the $v_i$-space, each choice of $\eta$ would be a point which lies in certain positive or negative region. To determine stability, it is equivalent to checking whether there are any points in the negative regions.

\begin{figure}[h]
	\centering
	\begin{subfigure}{5cm}
		\centering
		\includegraphics[width=5cm]{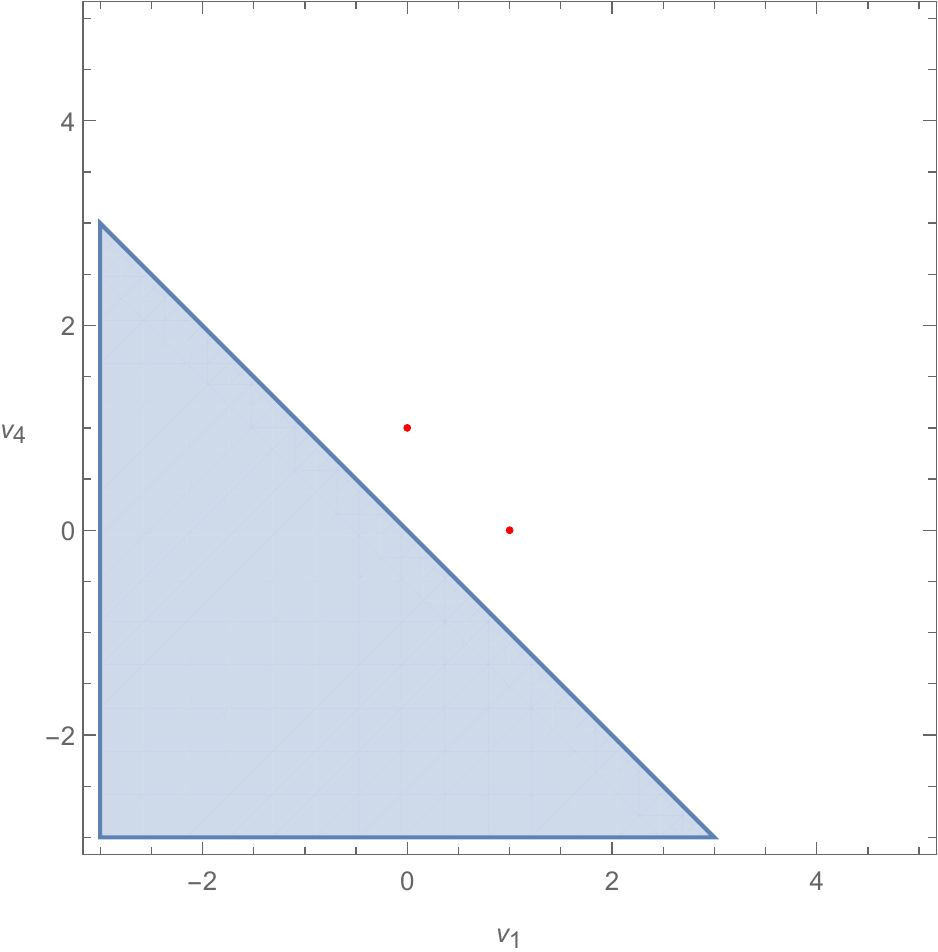}
		\caption{}
	\end{subfigure}
	\begin{subfigure}{5cm}
		\centering
		\includegraphics[width=5cm]{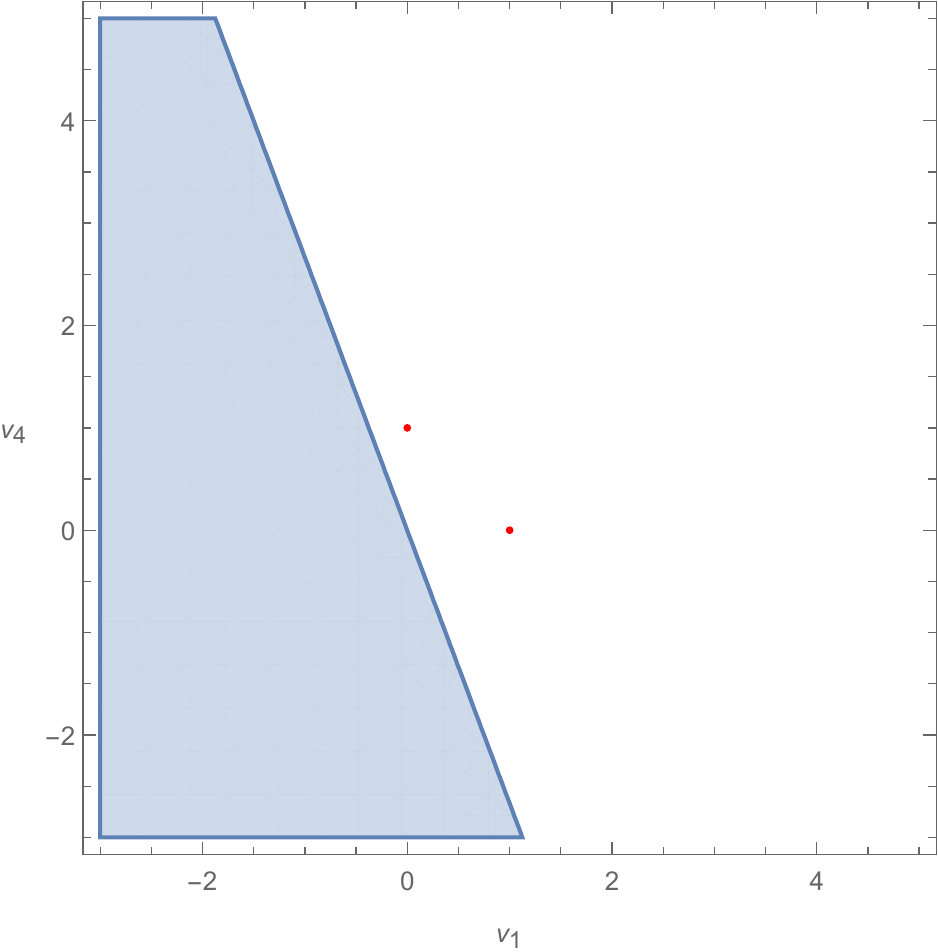}
		\caption{}
	\end{subfigure}
    \begin{subfigure}{5cm}
    	\centering
    	\includegraphics[width=5cm]{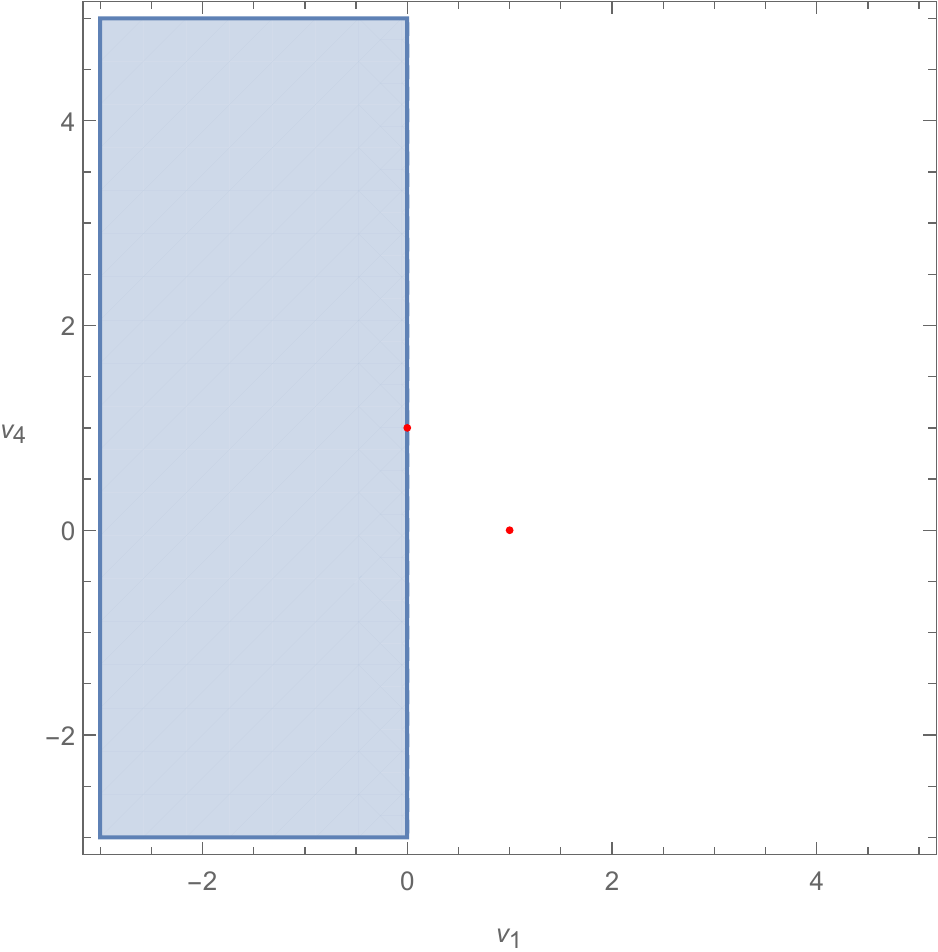}
    	\caption{}
    \end{subfigure}
    \begin{subfigure}{5cm}
    	\centering
    	\includegraphics[width=5cm]{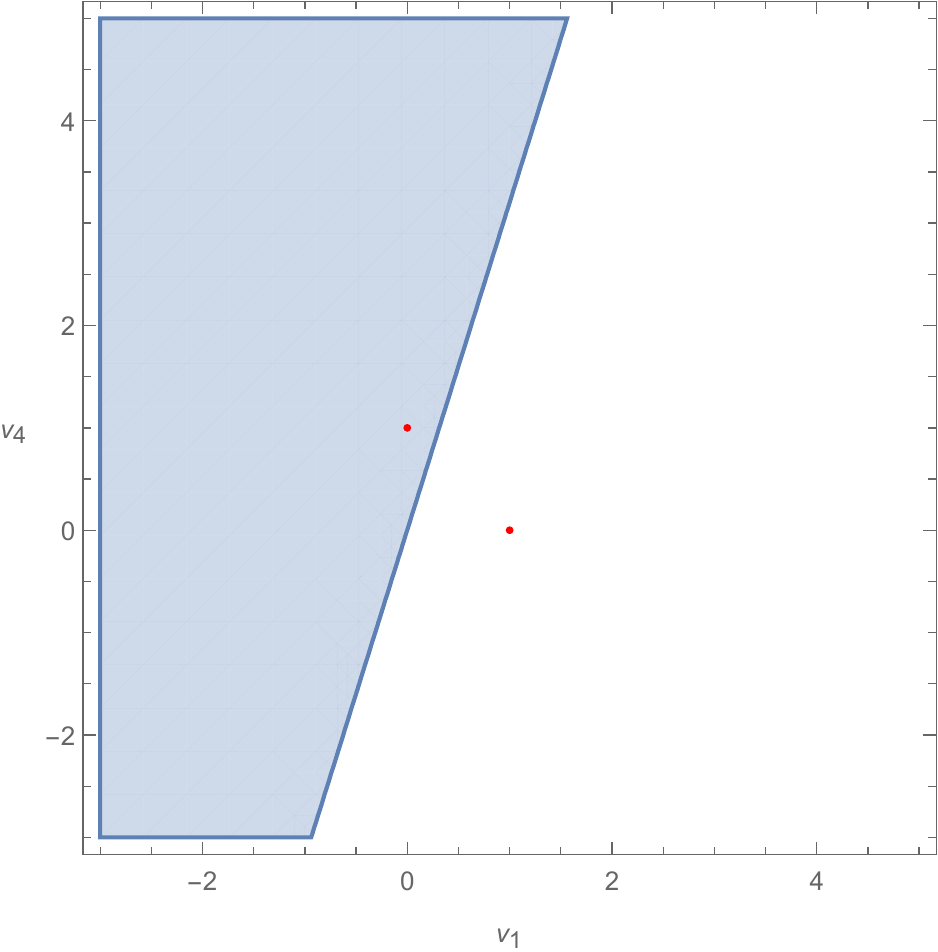}
    	\caption{}
    \end{subfigure}
    \begin{subfigure}{5cm}
    	\centering
    	\includegraphics[width=5cm]{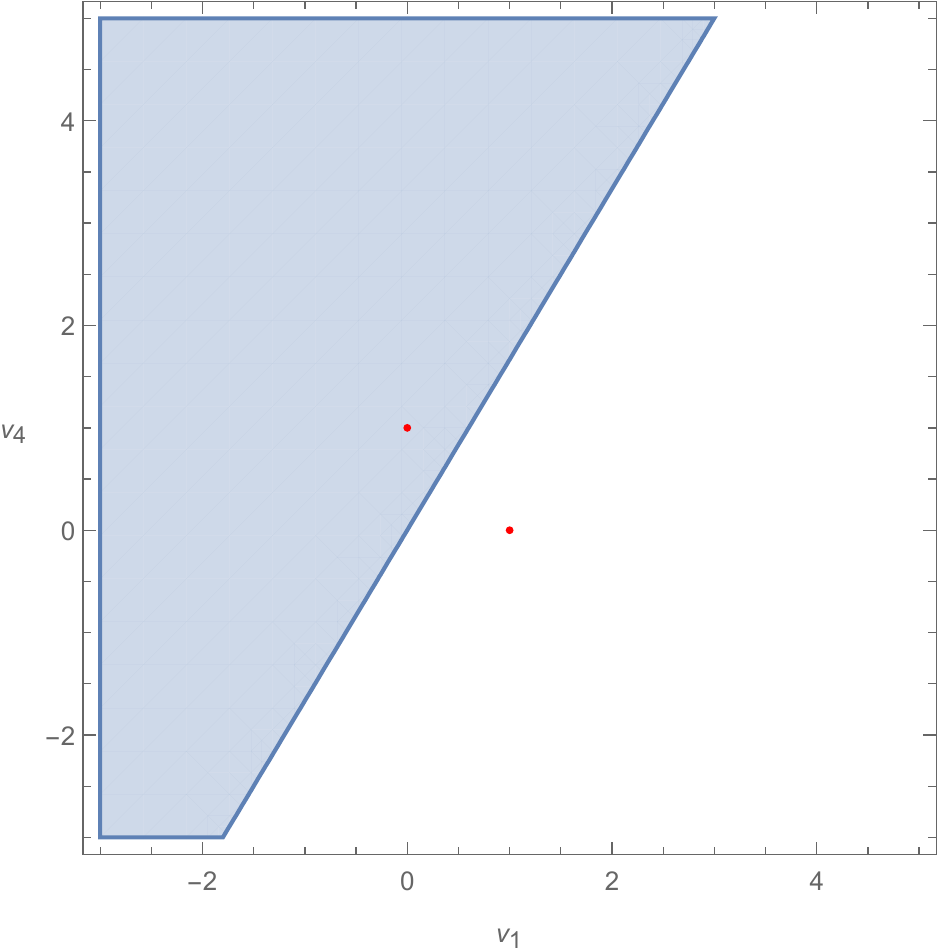}
    	\caption{}
    \end{subfigure}
    \begin{subfigure}{5cm}
    	\centering
    	\includegraphics[width=5cm]{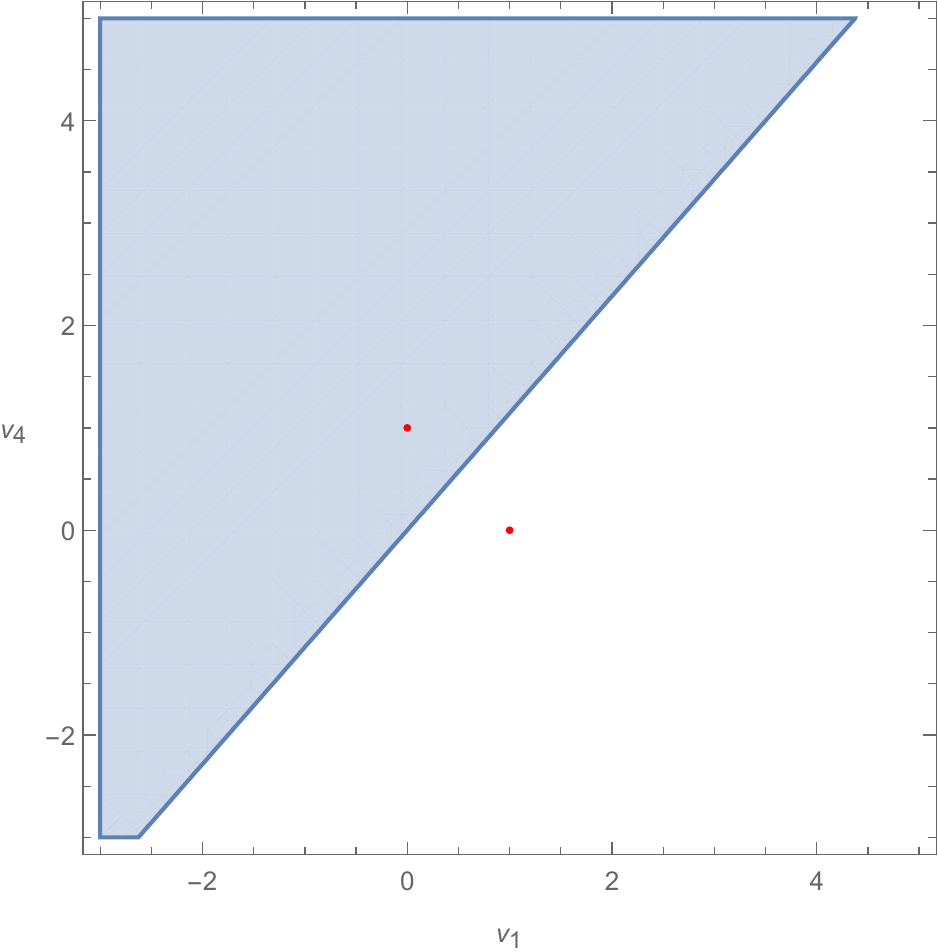}
    	\caption{}
    \end{subfigure}
	\caption{The region plot in $v_i$-space for $w^2+x^2+y^2+z^{n+1}=0$ at $n=1,\dots,6$ shown in (a)$\sim$(f) respectively. The red points correspond to the two test symmetries and the blue area is the region where $F<0$ in each picture. A red point inside the blue region indicates that the ring is unstable.}\label{A3folds}
\end{figure}

For example, consider the Futaki invariant for the hypersurface $w^2+x^2+y^2+z^{n+1}=0$ and test symmetry $\eta$ with charges $(v_1,v_2,v_3,v_4)$. Its Futaki invariant is given in \eqref{futex}. It is often difficult to visualize the $v_i$-space, but here since the coordinates $w,~x,~y$ are symmetric, we can solely consider $v_1$ and $v_4$ (i.e. two ways of dropping terms, although we can use some specific method to reduce the number of test symmetries in this case). We depict some $v_1$-$v_4$ planes for small $n$'s in Figure \ref{A3folds}.
Indeed, we see that the ring is only stable for $n=1,2$ as there is no red point inside the negative region which agrees with the result in \cite{Collins:2016icw}. We should be careful with $n=3$ where a red point lives on the boundary of the blue region, showing that $F=0$. The test configuration is certainly not trivial, and by computing the norm for this test symmetry with charges (0,0,0,1), we get $||\eta||^2=27/128\neq0$. Hence, the ring is unstable for $n=3$.

Some simplifications can be made to reduce the number of necessary test symmetries. In \cite{2015arXiv151207213C,Collins:2016icw}, $\eta$ is required to be normal and commuting with the automorphism group of $X$. For $X\subset\mathbb{C}^n$, the torus action and $\eta$ are induced by the subgroups of GL$(n,\mathbb{C})$. The commutation condition then implies that we can diagonlize the $T$- and $\mathbb{C}^*$-subgroups simultaneously. Hence, we will always assume that the test symmetries are diagonal under some unitary changes of basis. Normality could be boiled down to two conditions called Serre's criterion: $S_2$ and $R_1$. It is often not easy to check the former, but as we are always dealing with Cohen-Macaulay rings, $S_2$ is always guaranteed. Therefore, only $R_1$, namely being regular in codimension one, is left. This means that the singular locus has codimension no less than two, which can be checked via the Jacobian. We may also use \texttt{Macaulay2} \cite{M2} and the package \texttt{FastLinAlg} \cite{FastLinAlg} to tell this. In fact, we are also allowed to consider more general test configurations that are not normal or even those who have test symmetry not commuting with the $T$-action, but they will not give any additional information\footnote{The condition of being normal is related to the triviality of the central fibre. It was discussed in \cite{2011arXiv1111.5398L} that normality could avoid some pathological test configurations. However, as pointed out in \cite{szekelyhidi2014introduction}, we can instead use an alternative definition by introducing the norm whose vanishing is sufficient to give K-stability (when $F$ is zero). Regarding the norm, there could also be different conventions as aforementioned, and here we take the definition as in \eqref{norm}. We are grateful to G\'abor Sz\'ekelyhidi for clear explanations on this.}. For simplicity, we will therefore not require the normality condition as this should not affect our results.

\subsubsection{``Problematic'' Test Symmetries}\label{probs}
Following the above procedure to compute the Futaki invariant, especially using \eqref{Futaki4}, one can easily find some inconsistencies that seems to give ``sick'' test symmetries\footnote{As we will see, these $\eta$'s are not really ``problematic'' or ``sick''. We are just not using the correct way to do the computation.}.

\paragraph{The non-zero norm problem} The first problem is actually already resolved when defining the norm. Usually, a norm is defined only with the second line in \eqref{norm}, but we have to add the first line which makes the definition seemingly weird. For instance, for $\mathbb{C}^{3}=\text{Cone}\left(S^5\right)$ (or more generally, $\mathbb{C}^{n}=\text{Cone}\left(S^{2n-1}\right)$), there would be no non-trivial test configurations as this is toric with a maximal number of torus action. Indeed, we always have a vanishing Futaki invariant. Its stability is for sure expected as physically this corresponds to the $\mathcal{N}=4$ SYM in 4d which is superconformal. However, all the test symmetries, except the one with charge $(1,1,1)$, would yield non-zero norms. Another less ``trivial'' example is the conifold $uv+y^2+z^2=0$ and the test symmetry with charges $(1,-1,0,0)$ (though we would not have this if we make a linear holomorphic change to $w^2+x^2+y^2+z^2=0$), which leads to $F=0$. Such test configuration is certainly trivial, but $(c_0-b_0^2/a_0)=1/3\neq0$. However, the conifold is undoubtedly stable as it admits a Ricci-flat cone metric.

\paragraph{The $\epsilon$-region problem} Recall that physically we are only focusing on the $\epsilon>0$ region for $a_0(\zeta(\epsilon))$ to find whether there is a minimum because we want $\epsilon(\eta-a\zeta)$ to give the same central fibre as the test symmetry $\eta$ does. However, if we consider $w^2+x^2+y^2+z^5=0$ and $\eta$ with $(-1,-1,-1,0)$, we find that $(\eta-a\zeta)$ would give rise to $a=-3/2$ and weights $(8/7,8/7,8/7,6/7)$, which has an opposite central fibre. This seems to indicate that we should look at the region with $\epsilon<0$ in this case. Consequently, $F<0$ here would not destabilize the ring. However, we know from Figure \ref{A3folds} and also \S\ref{ADE} that $(0,0,0,1)$, which has an equivalent test configuration as $(-1,-1,-1,0)$, is the right test symmetry that destabilizes the ring.  This becomes a bigger issue if we consider stable rings or even non-complete intersections. For instance, consider the orbifold $\mathbb{C}^3/(\mathbb{Z}_4\times\mathbb{Z}_2)$ $(1,0,3)(0,1,1)$ whose relations are given in \cite{Hanany:2012hi}:
\begin{equation}
x_1x_3=x_2^2,~y_1y_2=x_3^2,
\end{equation}
where $x_1$ has order $4/3$ and $x_2$ has order 2 with the remaining three having order $8/3$. Its Gr\"obner basis is
\begin{equation}
x_3^2-y_1y_2,~x_1y_1y_2-x_2^2x_3,~x_1x_3-x_2^2.\label{orbgroeb}
\end{equation}
Since this is a toric variety, it should be K-stable. Let the test symmetry have charges $(0,0,-1,0,0)$. Then the test configuration reads
\begin{equation}
t^{-2}x_3^2-y_1y_2,~x_1y_1y_2-t^{-1}x_2^2x_3,~t^{-1}x_1x_3-x_2^2.
\end{equation}
However, with $a=-1/2$, $\epsilon(\eta-a\zeta)$ has charges $\epsilon(2/3,1,1/3,4/3,4/3)$. The test configuration is
\begin{equation}
t^{2\epsilon/3}x_3^2-t^{8\epsilon/3}y_1y_2,~t^{10\epsilon/3}x_1y_1y_2-t^{7\epsilon/3}x_2^2x_3,~t^{2\epsilon}x_1x_3-t^{2\epsilon}x_2^2.
\end{equation}
Now, no matter what value $\epsilon$ takes, the two central fibres will never be the same. We do not even know which region of $\epsilon$ to consider.

\begin{figure}[h]
	\centering
	\begin{subfigure}{7cm}
		\centering
		\includegraphics[width=6cm]{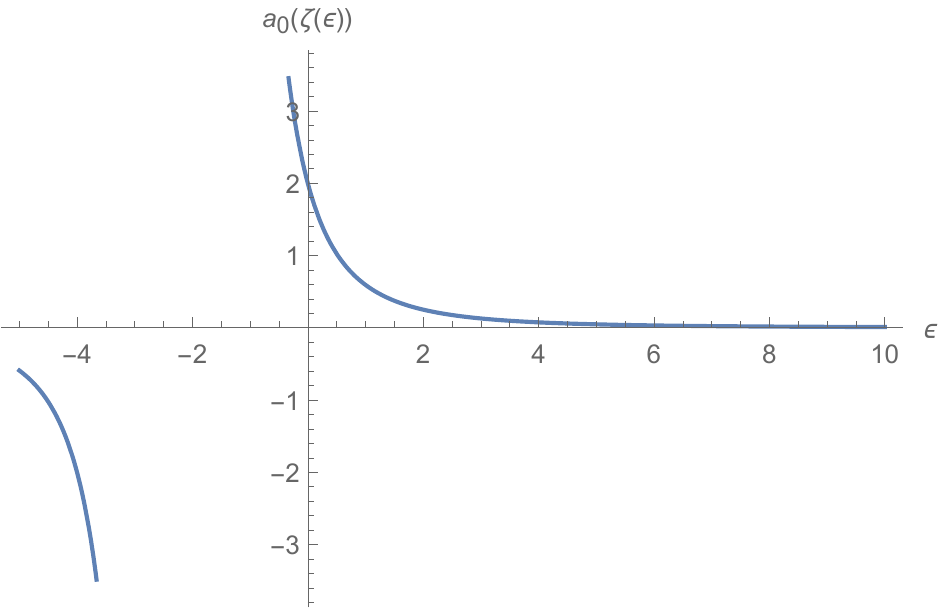}
		\caption{}
	\end{subfigure}
	\begin{subfigure}{7cm}
		\centering
		\includegraphics[width=6cm]{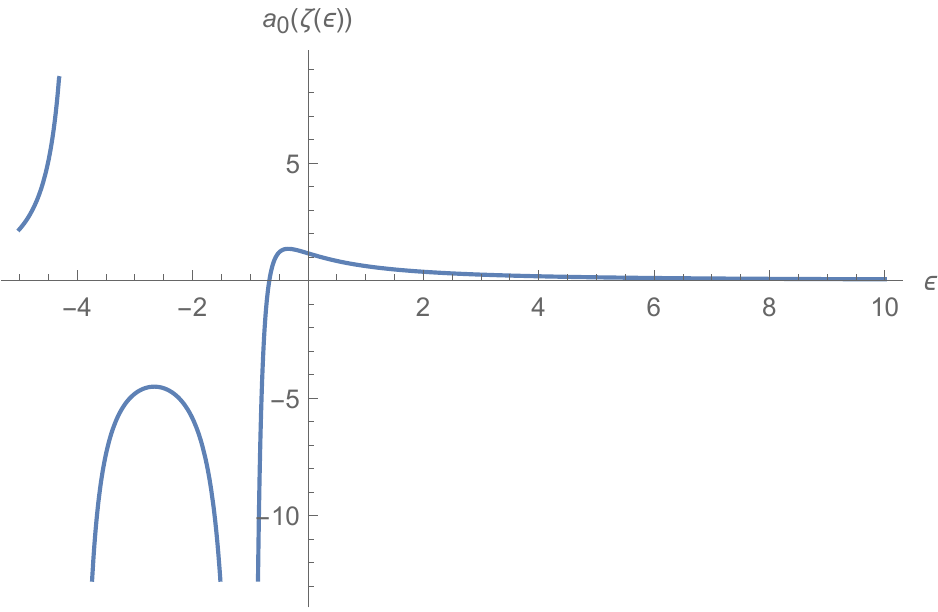}
		\caption{}
	\end{subfigure}
	\caption{(a) The conifold $w^2+x^2+y^2+z^2=0$ and $\eta$ with charges $(0,-1,-1,-1)$, and $a_0(\zeta(\epsilon))=\frac{16}{(\epsilon+2)^3}$. (b) The $E_7$ threefold $w^2+x^2+y^3+yz^3=0$ and $\eta$ with charges $(-1,-1,0,-1)$, and $a_0(\zeta(\epsilon))=\frac{750(3\epsilon+2)}{(\epsilon+4)(17\epsilon+18)^2}$.}\label{inf}
\end{figure}

\paragraph{The $F<0$ problem} Even if a test symmetry does not cause the $\epsilon$-region problem, the Futaki invariant we get could also be problematic. For example, let us consider the conifold $w^2+x^2+y^2+z^2=0$ and the test symmetry with charges $(-1,-1,-1,0)$. Now $a=-3/2$ and $(\eta-\epsilon\zeta)$ gives charges $(-5/2,-5/2,-5/2,-3/2)$. Therefore, we should still focus on the region of positive $\epsilon$. Following \eqref{Futaki4}, it is straightforward that $F=-3<0$. However, we already know that the conifold is stable. Under such construction, this contradiction can happen for any stable case. Another example is given in Figure \ref{inf}(b).

In the next subsection, we will see a method to resolve this, but if we insist on the results from \eqref{Futaki4}, we could physically understand the problem for a subset of these test symmetries. This can be explained if we contemplate the plots of $a_0(\zeta(\epsilon))$ against $\epsilon$ as in Figure \ref{inf}.
To destabilize the original ring, (the piece around some neighbourbood of $\epsilon=0$ of) the curve should have a local minimum at some positive $\epsilon$. However, the cases in Figure \ref{inf} do not have such local minima. In other words, $a_0(\zeta(\epsilon))$ keeps decreasing as $\epsilon\rightarrow\infty$, so if we consider the new R-symmetry parameterized by $\zeta(\epsilon)=\zeta+\epsilon(\eta-a\zeta)$, viz, $\zeta(\epsilon)/\epsilon=\zeta/\epsilon+(\eta-a\zeta)$ with $\epsilon\rightarrow\infty$, we would get $\eta=a\zeta$, which does not make sense. We should again emphasize that this could not account for all the ``sick'' $\eta$'s. For example, if we consider the test symmetry with $(-1,-1,-1,2)$ for the stable $A_2$ threefold $w^2+x^2+y^2+z^3=0$, then $a_0(\zeta(\epsilon))=\frac{375(\epsilon+2)}{16(3-\epsilon)^3(1+3\epsilon)}$. On the smooth piece around the neighbourhood of $\epsilon=0$, it has a local minimum at $\epsilon=(5\sqrt{10}-13)/9>0$.

\subsubsection{Regularizations of Numerators}\label{reg}
To find out what really goes wrong, it is always useful to start from the original definitions and derivations of K-stability. Recall that algebro-geometrically the Futaki invariant is defined as $F=B_0A_1/A_0-B_1$, where $A_i$'s and $B_i$'s are the leading and subleading coefficients of $d_k$ and $w_k$ respectively\footnote{Since the $a_i$'s have a different convention in this paper, we will use capital letters for the traditional conventions in mathematics literature such as \cite{szekelyhidi2014introduction,Collins:2012dh} to distinguish them.}. Therefore, we can compute $A_i$'s and $B_i$'s using their definitions and compare with the results from HS.

Let us again consider the conifold $w^2+x^2+y^2+z^2=0$. For the usual test symmetry $\eta(t)\cdot(w,x,y,z)=(w,x,y,tz)$, the central fibre is $w^2+x^2+y^2=0$, and the HS gives
\begin{equation}
	\text{HS}=\frac{1-t^2}{\left(1-t\right)^3\left(1-t^{1+\epsilon}\right)}=\frac{2}{(1+\epsilon)s^3}+\frac{2+\epsilon}{(1+\epsilon)s^2}+\dots\label{HSex}
\end{equation}
Taking $\epsilon=0$, we have (in the convention of \cite{Collins:2012dh})
\begin{equation}
	A_0(n-1)!=2A_0=2,~A_1(n-2)!=A_1=2.
\end{equation}
Likewise,
\begin{eqnarray}
	B_0&=&-\frac{1}{n}\text{D}_{\epsilon}A_0(\epsilon)|_{\epsilon=0}=-\frac{1}{3}\text{D}_{\epsilon}\left(\frac{1}{1+\epsilon}\right)\bigg|_{\epsilon=0}=\frac{1}{3},\nonumber\\
	B_1&=&-\frac{1}{n-1}\text{D}_{\epsilon}A_1(\epsilon)|_{\epsilon=0}=-\frac{1}{2}\text{D}_{\epsilon}\left(\frac{2+\epsilon}{1+\epsilon}\right)\bigg|_{\epsilon=0}=\frac{1}{2}.
\end{eqnarray}
By definition, the dimension of the degree $k$ piece $R_k$ of the ring is
\begin{equation}\label{eqn:dk}
	d_k=\binom{k+3-1}{k}+\binom{k-1+3-1}{k-1}=k^2+2k+1,
\end{equation}
where we have used that the number of independent monomials of degree $k$ with $m$ variables is $\binom{k+m-1}{k}$. In particular, the first term in \eqref{eqn:dk} counts the number of independent monomials of the form $x^ly^mz^p$ with $l+m+p=k$, while the second term counts the monomials of the form $wx^ly^mz^p$ with $l+m+p=k-1$.  Likewise, by definition, the total weight of $R_k$ is
\begin{equation}
	w_k=\sum_{i=0}^k\binom{k-i+2-1}{k-i}i+\sum_{i=0}^{k-1}\binom{k-i-1+2-1}{k-i-1}i=\frac{1}{3}k^3+\frac{1}{2}k^2+\frac{1}{6}k.
\end{equation}
Here, we see that the first term sums up the different choices for monomials weighted $i$ of the form $x^ly^mz^i$ with $l+m=k-i$, while the second term sums for monomials of the form $wx^ly^mz^i$ with $l+m=k-i-1$.
As we can see, the result from HS agrees with the one from definition for this test symmetry.

However, if we consider $\eta(t)\cdot(w,x,y,z)=\left(t^{-1}w,t^{-1}x,t^{-1}y,z\right)$, which yields $F<0$, the $A_i$'s remain the same while from
\begin{equation}
	\text{HS}=\frac{1-t^2}{\left(1-t\right)\left(1-t^{1-\epsilon}\right)^3}=-\frac{2}{(-1+\epsilon)^3s^3}+\frac{-2+3\epsilon}{(-1+\epsilon)^3s^2}+\dots,
\end{equation}
we get $B_0=-1$ and $B_1=-3/2$. On the other hand, by definition of $w_k$, $B_0=-2/3$ and $B_1=-3/2$. We see that the results are different.

Even for some non-negative $F$'s, we would still have this issue. Consider the test symmetry $\eta(t)\cdot(w,x,y,z)=\left(tw,tx,ty,tz\right)$. Then from HS, we have
\begin{equation}
	A_0=1,~A_1=2,~B_0=\frac{4}{3},~B_1=2,~F=\frac{2}{3}.
\end{equation}
In contrast, from definition, as $w_k$ is simply $kd_k$ here, we can easily get 
\begin{equation}
A_0=1,~A_1=2,~B_0=1,~B_1=2,~F=0.
\end{equation}
In fact, we expect the Futaki invariant to vanish for this test symmetry not only because this is the result from the computation using definition, but also because the test configuration $t^2\left(w^2+x^2+y^2+z^2\right)$ is trivial.

One may wonder if this is a matter of convention. In other words, it might be possible that we have not found the right convention that makes all the parameters agree. After all, the precise values can differ by a positive numerical factor in different conventions. This possibility can be excluded by the example $xz-y^2=0$ with $\eta(t)\cdot(x,y,z)=\left(t^{-1}x,ty,z\right)$. The HS is
\begin{equation}
	\text{HS}=\frac{1-t^{2p}}{\left(1-t^{p+\epsilon}\right)\left(1-t^{p-\epsilon}\right)\left(1-t^p\right)},
\end{equation}
where the convention is arbitrary with some power $p$. 
From this HS, we find that
\begin{equation}
	B_0=B_1=0.
\end{equation}
However, the correct answers are already obtained in \cite{szekelyhidi2014introduction} by definition:
\begin{equation}
	B_0=B_1=\frac{1}{2}.
\end{equation}
Hence, no matter what positive constant we multiply, the two would never agree. This shows that the problem is from the steps in the HS method we use.

In \cite{Collins:2012dh}, the index character is defined to be
\begin{equation}
	G(\zeta)=\sum\limits_{\alpha\in\mathfrak{t}^*}\text{e}^{-s\alpha(\zeta)}\dim R_\alpha,
\end{equation}
where $\mathfrak{t}$ is the Lie algebra of the torus action and $R_\alpha$ is the associated root space with root $\alpha$ in the root space decomposition of $R$. Since $\zeta\in\mathfrak{t}$ is a symmetry acting with positive weights, viz, a \emph{Reeb vector field}, the sum converges for Re$(s)>0$ and has a meromorphic extension at $s=0$. It is proven that the index character has a Laurent expansion
\begin{equation}
	G=\frac{A_0(n-1)!}{s^n}+\frac{A_1(n-2)!}{s^{n-1}}+\dots
\end{equation}
at $s=0$, which is exactly the HS. Similarly, to show that $b_i$'s (and also $c_0$) are certain derivatives of $a_i$'s, the weight character is defined to be
\begin{equation}
	C_\eta=\sum\limits_{\alpha\in\mathfrak{t}^*}\text{e}^{-s\alpha(\zeta)}\alpha(\eta)\dim R_\alpha.
\end{equation}
Then one can show that
\begin{equation}
	-tC_\eta=\frac{\partial}{\partial\epsilon}G(\zeta+\epsilon\eta)\bigg|_{\epsilon=0}.
\end{equation}
Importantly, this expression is true because for sufficiently small $\epsilon$, $(\zeta+\epsilon\eta)$ is a Reeb field, and hence the sum for $G(\zeta+\epsilon\eta)$ converges uniformly for $s>0$. Therefore, since the Reeb field determines the weights of the relations and the information of these relations are contained in the numerator of HS, we should modify the HS with $\epsilon$. In other words, we should also write the numerator with respect to the Reeb field $(\zeta+\epsilon\eta)$, rather than just $\zeta$.

When we write HS, we still need to consider $(\zeta+\epsilon\eta)$ as two degrees for the grading: one variable $t_0$ for $\zeta$ and one variable $t_1$ for $\eta$. Only after this step, we can assign small $\epsilon$ to the powers of $t_1$. However, in the first step, $\eta$ in fact is not a Reeb field and it would make the equations in the ideal inhomogenous. Therefore, we cannot simply write down the HS. One may try some homogenization of the equations, but it would not yield correct results for K-stability.

Here, we discuss a method to modify the numerator with the help of Gr\"obner basis\footnote{To the authors' best knowledge, such method has never been mentioned in literature. Modifying the numerators might be known to mathematics society, but mathematicians mainly focus on the aforementioned complexity one varieties (such as those in \S\ref{ADE} below), where one only needs to check several test symmetries using the method in \cite{2015arXiv150704442I}. It turns out that the remaining possible test symmetries are simple enough so that no modifications of numerators are required. The authors also consulted some mathematicians, but modifying the numerators was never mentioned. Therefore, it is worth spelling out such method here. Any comment on this is more than welcomed.}. As discussed in Appendix \ref{hsrev}, when writing HS, it suffices to consider the initial terms of the equations in the Gr\"obner basis. In particular, the initial terms are obtained from some ordering of the variables, and likewise, the initial terms for the flat limit are also obtained from a specific ordering, that is, the (lowest) powers/weights of $t$ in the relations\footnote{We are using $t$ both in the HS and in the test configuration, but it should be clear which $t$ we are referring to in the context.}.

Therefore, to write the HS with respect to $(\zeta+\epsilon\eta)$, especially the $t_1$ for $\eta$, we also take the initial terms induced by the same ordering when taking flat limits. If the initial term has a factor $t^p$ (regardless of the sign of $p$), then we should include the corresponding power of $t_1$ in the numerator. If the initial term has no $t$, then the numerator is free of $t_1$. In particular, the power of $t_1$ is determined by the power of initial terms of the ideal. 

For instance, for the conifold example above, $(0,0,0,1)$ would still give the same HS as in \eqref{HSex}. For $(-1,-1,-1,0)$, the initial term would have $t^{-1}$, and therefore we should add some power of $\epsilon$ in the numerator. We see that the ideal of conifold is quadratic, so we add a factor of $t_1^{-2\epsilon}$ to the numerator. Let $t_0$ and $t_1$ denote the variables for $\zeta$ and $\epsilon\eta$ respectively. The multivariate (refined) HS reads
\begin{equation}
	\text{HS}=\frac{1-t_0^2t_1^{-2\epsilon}}{\left(1-t_0\right)\left(1-t_0t_1^{-\epsilon}\right)^3}.
\end{equation}
Unrefining the HS by $t_0=t_1=t$, we get
\begin{equation}
	\text{HS}=\frac{1-t^{2(1-\epsilon)}}{\left(1-t\right)\left(1-t^{1-\epsilon}\right)^3}.
\end{equation}
From this HS, following the usual steps of taking Laurent series and derivatives, we find that
\begin{equation}
	A_0=1,~A_1=2,~B_0=-\frac{2}{3},~B_1=-\frac{3}{2},
\end{equation}
which is exactly the same result obtained from definition. Indeed, this yields $F=1/6>0$, which equals to the Futaki invariant for $(0,0,0,1)$. This agrees with the fact that the two test symmetries give rise to equivalent test configurations\footnote{Notice that in our convention where $a_0=a_1$, the value of Futaki invariant has an extra dimensional factor $n(n-1)$. For example, here we have $a_1b_0/a_0-a_1=3\times(3-1)\times1/6=1$.}.

We may also check that for $(1,1,1,0)$,
\begin{equation}
	\text{HS}=\frac{1-t^{2}}{\left(1-t\right)\left(1-t^{1+\epsilon}\right)^3}
\end{equation}
since the initial term is $z^2$ which has weight $t^0$, and that for $(0,0,0,-1)$,
\begin{equation}
	\text{HS}=\frac{1-t^{2(1-\epsilon)}}{\left(1-t^{1-\epsilon}\right)\left(1-t\right)^3}
\end{equation}
since the initial term is $t^{-2}z^2$. Again, we can verify that both of them yield the same correct $A_i$'s and $B_i$'s as those from definition, as well as a positive Futaki invariant. Likewise, one can also check that the $xz-y^2$ example gives the correct $B_0=B_1=1/2$.

We can also verify that by modifying the numerators, for the aforementioned problems, we would not have the $\epsilon$-region issue or negative $F$ for stable rings any more. We will omit the detailed calculations here. Nevertheless, it is worth noting that some trivial test configurations will thence automatically have $F=0$ and even a vanishing norm. Recall that without the modification of numerators, $(1,1,1,1)$ yields a positive $F$, as well as a non-zero norm. After regularizing the numerator,
\begin{equation}
	\text{HS}=\frac{1-t^{2(1+\epsilon)}}{\left(1-t^{1+\epsilon}\right)^4}.
\end{equation}
This gives the correct $A_0=B_0=C_0=1$ and $A_1=B_1=2$. Thus, $F=A_1B_0/A_0-B_1=0$ and $||\eta||^2=C_0-B_0^2/A_0=0$ as expected.

However, we still need the first line in the definition \eqref{norm} of the norm. For example, when we write the conifold as $uv=xy$, and consider $(1,-1,0,0)$, the numerator still remains the same. Hence, $C_0-B_0^2/A_0$ is still not zero. However, such test symmetry is a bit special and we can still force the norm to vanish via definition. Incidentally, we find that if the HS is written as
\begin{equation}
	\text{HS}=\frac{1-t^{2\left(1-\epsilon^2\right)}}{\left(1-t^{1-\epsilon}\right)\left(1-t^{1+\epsilon}\right)\left(1-t\right)^2},
\end{equation}
then $C_0-B_0^2/A_0=0$. Similarly, for $(1,-1,1,-1)$, if we write the HS as
\begin{equation}
\text{HS}=\frac{1-t^{2\left(1-\epsilon^2\right)}}{\left(1-t^{1-\epsilon}\right)^2\left(1-t^{1+\epsilon}\right)^2},
\end{equation}
then $C_0-B_0^2/A_0=0$ as well. So far it is still not clear why this happens. It might be possible that it requires higher order of corrections in the numerator for such special test symmetries, or maybe this is just a coincidence.

Now in our convention with $a_i$ and $b_i$, although they take values different from those obtained by definition. They would always differ by a positive constant depending only on dimension, viz,
\begin{equation}
	\frac{a_1}{a_0}b_0-b_1=n(n-1)\left(\frac{A_1}{A_0}B_0-B_1\right).
\end{equation}
The norms (squared) agree up to the same positive constant as well. Therefore, this method can certainly be applied in any convention.

\subsubsection{The Rescaling Method}\label{rescaling}
We now have seen how to write the HS and get the Futaki invariants correctly by some modifications in the numerators. However, in principle, there could be a large number of possible test symmetries to determine K-stability and such method does not reduce this number. Here, by considering the central fibres, we propose a method that potentially simplifies the process of checking test symmetries.

In general, if the test symmetry has charge $(v_1,\dots,v_m)$, then the test configuration for $I=\langle f_1,\dots,f_l\rangle$ is generated by $f_1\left(t^{v_1}x_1,\dots,t^{v_m}x_m\right),\dots,f_l\left(t^{v_1}x_1,\dots,t^{v_m}x_m\right)$. When taking the flat limit, only the initial terms would survive as discussed in \S\ref{flat}. Another way to view the flat limit is by considering a rescaling of the $f_i$'s \cite{1993alg.geom..4003B}. Under the rescaling, we write $g_1=t^{w_1}f_1\left(t^{v_1}x_1,\dots,t^{v_m}x_m\right),\dots,g_l=t^{w_l}f_l\left(t^{v_1}x_1,\dots,t^{v_m}x_m\right)$ such that the initial terms in each $f_i$ has weight zero with respect to $t$. Then at $t=1$, we recover $I=\langle g_1,\dots,g_l\rangle|_{t=1}$, and at $t=0$, we recover the flat limit $I_0=\langle g_1,\dots,g_l\rangle|_{t=0}$. For example, $(0,-1,-1,-1)$, which has $F<0$ by \eqref{Futaki4} without regularizing the numerator, gives $f=w^2+t^{-2}x^2+t^{-2}y^2+t^{-2}z^2$ for the conifold, and we can rescale it to $g=t^2f=t^2w^2+x^2+y^2+z^2$. It is worth noting that this $g$ is what we get directly from $(1,0,0,0)$ without rescaling. We may also consider $(-1,-1,-1,-1)$ which gives negative Futaki invariant if we naively use \eqref{Futaki4} to do the calculation. However, $t^{-2}w^2+t^{-2}x^2+t^{-2}y^2+t^{-2}z^2$ is simply a trivial test configuration and can be rescaled to $w^2+x^2+y^2+z^2$. Indeed, we would just get the trivial $\eta'=0$.
 
Inspired by this, suppose we pick a test symmetry $\eta$ with a random charge, then we may follow these steps to only compute $F$ for $\eta'$:
\begin{itemize}
	\item We rescale the $f_i$'s to $g_i$'s such that the terms with lowest $t$-weights would have weight 0. This would lead to some new test symmetry $\eta'$ that directly yields $g_i$'s without any rescaling. Since all the initial terms have no $t$'s and no regularization in the numerator is required, we can simply use \eqref{Futaki4} to compute the Futaki invariant.
	\item When dealing with non-hypersurfaces, it is possible to have some $\eta$ whose rescaling (though we can always do such rescaling) does not correpsond to any $\eta'$. In other words, such configuration cannot have a test symmetry with all the initial terms having weight 0 for all the equations. In this case, we should find a ``minimal'' $\eta'$ in the sense that the number of $g_i's$ with non-zero lowest weights is minimized. Moreover, these non-zero lowest weights should be positive. In this situation, there is at least one initial term having a positive $t$-weight. Therefore, we should apply the modification of the numerator to compute $F$.
\end{itemize}

At the first step, we have already seen such examples as those for the conifold. It is easy to check that this also works for positive Futaki invariants. For instance, $(1,1,1,1)$ for the conifold can be rescaled to $(0,0,0,0)$ as well, both of which have trivial test configuration. Moreover, for those like $(1,-1,0,0)$ for $uv=xy$ which does not receive regularization in the numerator but with $c_0-b_0^2/a_0\neq0$, we can also rescale it to the trivial test configuration. Let us now contemplate some less non-trivial example whose K-stability is known to validate this. Consider the aforementioned orbifold $\mathbb{C}^3/(\mathbb{Z}_4\times\mathbb{Z}_2)$ (1,0,3)(0,1,1) with $\eta$-charges $(1,1/2,-1/2,-1,0)$ whose test configuration is
\begin{equation}
	t^{-1}x_3^2-t^{-1}y_1y_2,~x_1y_1y_2-t^{1/2}x_2^2x_3,~t^{1/2}x_1x_3-tx_2^2,
\end{equation}
which should be rescaled according to the above steps. Indeed, a naive computation for this yields a negative $F$. Then the test configuration can be written as
\begin{equation}
	x_3^2-y_1y_2,~x_1y_1y_2-t^2x_2^2x_3,~x_1x_3-t^2x_2^2
\end{equation}
with $\eta'$ giving charges $(0,1,0,0,0)$. We can then simply apply \eqref{Futaki4} which yields a positive $F$.

For the second step, let us consider the same orbifold with test charges $(0,-1,0,-1,-1)$ whose test configuration is
\begin{equation}
x_3^2-t^{-2}y_1y_2,~t^{-2}x_1y_1y_2-t^{-2}x_2^2x_3,~x_1x_3-t^{-2}x_2^2.
\label{eq:C3/Z4_ideal}
\end{equation}
Under the rescaling, the test configuration can be written as
\begin{equation}
t^2x_3^2-y_1y_2,~tx_1y_1y_2-tx_2^2x_3,~t^2x_1x_3-x_2^2\label{orbtest}
\end{equation}
with $\eta'$ giving charges $(1,0,1,0,0)$. Note that we can not simply rescale every relation in the ideal such that the initial term has weight 0 in $t$. For example, the first and third relations in \eqref{eq:C3/Z4_ideal} show that $x_3$ should have non-trivial weight and $x_2$ should have weight 0. This then fixes the form of the second relation to be that shown in \eqref{orbtest}. It turns out that for $\eta'$
\begin{equation}
	\text{HS}=\frac{1-t^4-t^{16/3}+t^{28/3}}{\left(1-t^{4/3+\epsilon}\right)\left(1-t^2\right)\left(1-t^{8/3+\epsilon}\right)\left(1-t^{8/3}\right)^2},
\end{equation}
where it has no $\epsilon$'s in the numerator, and we can therefore use \eqref{Futaki4} to get $F>0$. However, we will see in \S\ref{ewsm}, in general there could be modifications in the numerator for $\eta'$ in the second step.

It is also possible that for $\eta_1$ and $\eta_2$ with different $\eta_1'$ and $\eta_2'$ have the same central fibre, but they are not related by a simple rescaling. For hypersurfaces, these are often equivalent as $\eta_2'=s\eta_1'$ for $s>0$ such as $(1,0,0,0)$ and $(2,0,0,0)$ for the aforementioned conifold example. Therefore, it suffices to consider only one of them. More generally, including non-hypersurfaces, it would be natural to speculate that $\eta_1'$ and $\eta_2'$ also give the same result as they lead to the same central fibre. Suppose we have $m$ monomials in all the equations, then there would be at most $(2^m-2)$ ways to drop terms (excluding dropping all terms or dropping no terms). This gives finitely many test symmetries although the number increases drastically when $m$ increases and this does not tell us the exact (minimal) number of test symmetries or exactly which test symmetries we need to check (compared to complexity one varieties in \cite{2015arXiv150704442I}). The above steps are based on the following point, which is yet to manifest. Using rescaling, we are actually choosing a representative for each central fibre, so either the representative test symmetry should be able to correctly indicate whether the variety can be destabilized to the central fibre, or maybe every test symmetry with the same central fibre should give the same sign of $F$.

In fact, a consequence of such rescaling is that there are only two ways to get a negative $F$. One possibility is that the $\zeta$-weight $k$ of a generator is small enough so that $nk-2<0$ in \eqref{Futaki4}, such as the A-type threefolds in \S\ref{ADE} below. The other possibility is that we have some negative weight in $\eta$, but this negative power of $t$ gets cancelled by other positive powers in the monomials in the relations. Then if the generator with this negative $\eta$-weight has a large enough $k$, the Futaki invariant could become negative. Such example includes the D-type threefolds in \S\ref{ADE} below.

These two ways of destabilizing the chiral ring should have explanations in terms of the dynamics of physics. The first way could be caused by the violation of unitarity bound. In particular, if a generator violates the unitarity bound, we would have $k<2/3$, which is exactly $3k-2<0$ from \eqref{Futaki4} for a three dimensional moduli space, such as the case for D3-branes probing CY$_3$. For higher dimensional moduli spaces, as we will see in \S\ref{examples}, the orders $k$ are not necessarily equal to R-charges numerically in the convention of $a_0=a_1$, and more importantly, it could be possible that (violation of) the unitarity bound ``leaks'' out of the $nk-2<0$ region (see for example \S\ref{sqcd}). For the first way, being unstable could also be caused by irrelevance of superpotential terms or some unknown dynamical reasons. For the second way, as shown in \cite{Collins:2016icw}, there could also be some unknown dynamical effects to prevent the ring from being a ring for an SCFT, such as the D-type threefolds.

\section{Illustrative Examples}\label{examples}
Now let us contemplate various examples to illustrate the above discussion. We will see \eqref{Futaki4} and the modification of numerator applied to different cases including non-complete intersections, and also how the rescaling method might reduce the number of possible test symmetries for equations whose variables have certain symmetries.

\subsection{ADE Threefolds}\label{ADE}
The Kleinian singularities can be obtained by orbifolding $\mathbb{C}^2$ with some subgroups $\Gamma$ of $SU(2)$, which are related to (affine) ADE Dynkin diagrams by McKay correspondence \cite{mckay}. We may require $a_0=a_1$ so that the canonical (2,0)-form has charge 2. However, they should always be stable as there would be no normal central fibres (and non-normal ones would not give any extra information). Hence, we can lift the ADE singularities to ``ADE threefolds'' \cite{Fazzi:2019gvt} by adding another squared term of a new coordinate to the defining equation\footnote{Note that these ADE threefolds are not to be confused with $\mathbb{C} \times \mathbb{C}^2 / \Gamma$ which are extensively used in D-brane quiver gauge theories, whose chiral rings are all stable.}. As one may check, the stabilities should be consistent with the results in \cite{2015arXiv151207213C,Fazzi:2019gvt}.

\paragraph{Cyclic group $\mathbb{Z}_{n+1}$: $\hat{A}_{n}$} The defining equation is $w^2+x^2+y^2+z^{n+1}=0$. This belongs to the family of Brieskorn-Pham (BP) singularity, also known as the Yau-Yu singularity of type I (YY-I) \cite{Collins:2012dh}. This ring $X$ has a symmetry $\zeta$ with charges $\left(\frac{2n+2}{n+3},\frac{2n+2}{n+3},\frac{2n+2}{n+3},\frac{4}{n+3}\right)$. Hence, we write the HS as
\begin{equation}
	\text{HS}=\frac{1-t^{(4n+4)/(n+3)}}{\left(1-t^{4/(n+3)}\right)\left(1-t^{(2n+2)/(n+3)}\right)^3}.
\end{equation}
Under Laurent expansion around $s=0$, we obtain $a_0(\zeta)=a_1(\zeta)=\frac{(n+3)^3}{8(n+1)^2}$. By \eqref{Futaki4},
\begin{equation}
	F=(v_1+v_2+v_3)\frac{n(n+3)^3}{8(n+1)^3}+v_4\frac{(3-n)(n+3)^3}{32(n+1)^2}\label{futex}
\end{equation}
for test symmetry with charges $(v_1,v_2,v_3,v_4)$. It suffices to check the test symmetries $\eta_i$ with charge $\delta_{ij}$ on the $j^\text{th}$ coordinate. In particular, (0,0,0,1) gives us the non-trivial result: $0<n<3$ \footnote{As aforementioned in Figure \ref{A3folds}, when $n=3$, the Futaki invariant is zero, but it is unstable since $||\eta||\neq0$. Also, if the $v_i$'s are complicated, we should modify the numerator to get the correct Futaki invariant rather than directly apply \eqref{Futaki4}. However, for hypersurfaces, they can all be rescaled such that the lowest $t$-weights are 0 in the equation. We will not restate these two points for similar situations below.} for K-stability.

\paragraph{Dicyclic group Dic$_{n-1}$: $\hat{D}_{n+1}$ ($n\geq3$)} The defining equation is $w^2+x^2+y^2z+z^n=0$. This belongs to the singularity of type YY-II. The ring $X$ has a symmetry $\zeta$ with charges $\left(\frac{2n}{n+1},\frac{2n}{n+1},\frac{2n-2}{n+1},\frac{4}{n+1}\right)$. Hence, we write the HS as
\begin{equation}
\text{HS}=\frac{1-t^{4n/(n+1)}}{\left(1-t^{4/(n+1)}\right)\left(1-t^{(2n-2)/(n+1)}\right)\left(1-t^{2n/(n+1)}\right)^2}.
\end{equation}
Under Laurent expansion around $s=0$, we obtain $a_0(\zeta)=a_1(\zeta)=\frac{(n+1)^3}{8n(n-1)}$. By \eqref{Futaki4},
\begin{equation}
F=(v_1+v_2)\frac{(n+1)^3(2n-1)}{16n^2(n-1)}+v_3\frac{(n+1)^3(n-2)}{8n(n-1)^2}+v_4\frac{(n+1)^3(5-n)}{32n(n-1)}
\end{equation}
for test symmetry with charges $(v_1,v_2,v_3,v_4)$. It suffices to check test symmetries $(0,0,-1/2,1)$, which yields
\begin{equation}
F=-\frac{(n+1)^3(n-2)}{16n(n-1)^2}+\frac{(n+1)^3(5-n)}{32n(n-1)}=-\frac{(n+1)^3(n^2-4n+1)}{32n(n-1)^2}.
\end{equation}
In addition, for test symmetries $(1,0,0,0)$ and $(0,0,1,0)$, we see that $F>0$ for $n>3$. Hence, the ring is stable when $n^2-4n+1<0$. Therefore, only the ring of $\hat{D}_4$ with $n=3$ is stable.

\paragraph{Binary tetrahedral/icosahedral group $\mathbb{BT}$, $\mathbb{BI}$: $\hat{E}_{6,8}$} The defining equation is $w^2+x^2+y^3+z^n=0$, where $n=4$ for $\mathbb{BT}$ and $n=5$ for $\mathbb{BI}$. This belongs to the singularity of type YY-I. The ring has a symmetry $\zeta$ with charges $\left(\frac{3n}{3+n},\frac{3n}{3+n},\frac{2n}{3+n},\frac{6}{3+n}\right)$. Hence, we write the HS as
\begin{equation}
\text{HS}=\frac{1-t^{6n/(3+n)}}{\left(1-t^{3n/(3+n)}\right)^2\left(1-t^{2n/(3+n)}\right)\left(1-t^{6/(3+n)}\right)}.
\end{equation}
Under Laurent expansion around $s=0$, we obtain $a_0(\zeta)=a_1(\zeta)=\frac{(n+3)^3}{18n^2}$. By \eqref{Futaki4},
\begin{equation}
F=(v_1+v_2)\frac{(n+3)^3(7n-6)}{108n^3}+v_3\frac{(n+3)^3(2n-3)}{36n^3}+v_4\frac{(n+3)^3(6-n)}{108n^2}
\end{equation}
for test symmetry with charges $(v_1,v_2,v_3,v_4)$. It suffices to check test symmetry $(0,0,0,1)$, and hence the ring is stable when $2\leq n<6$, in particular for $n=4,5$ here. For other test symmetries, $(1,0,0,0)$ and $(0,0,1,0)$, we see that $F>0$ since $n\geq2$.

\paragraph{Binary octahedral group $\mathbb{BO}$: $\hat{E}_{7}$} The defining equation is $w^2+x^2+y^3+yz^3=0$. This belongs to the singularity of type YY-II. The ring has a symmetry $\zeta$ with charges $\left(\frac{9}{5},\frac{9}{5},\frac{6}{5},\frac{4}{5}\right)$. Hence, we write the HS as
\begin{equation}
\text{HS}=\frac{1-t^{18/5}}{\left(1-t^{9/5}\right)^2\left(1-t^{6/5}\right)\left(1-t^{4/5}\right)}.
\end{equation}
Under Laurent expansion around $s=0$, we obtain $a_0(\zeta)=a_1(\zeta)=\frac{125}{108}$. By \eqref{Futaki4},
\begin{equation}
F=\frac{2155}{1944}(v_1+v_2)+\frac{125}{162}v_3+\frac{125}{432}v_4
\end{equation}
for test symmetry with charges $(v_1,v_2,v_3,v_4)$. It suffices to check test symmetries $(1,0,0,0)$, $(0,0,1,-1/3)$ and $(0,0,0,1)$, and hence the ring is stable. In \cite{Fazzi:2019gvt}, it was shown that the $E_7$ threefold does not admit a \emph{non-commutative crepant resolution} (NCCR). Therefore, it is still possible to be an SCFT, but it could not have a string embedding. In other words, in light of Conjecture \ref{conj1}, this could be an SCFT without a D-brane system picture\footnote{It is also suggested that this could be a non-Lagrangian theory. We would like to thank Alessandro Tomasiello for pointing this out.}.

\subsection{del Pezzo Spaces}\label{dP}
Let us consider the del Pezzo family dP$_n$ where $0\leq n\leq8$. The HS is \cite{Benvenuti:2006qr}
\begin{equation}
	\text{HS}=\frac{1+(7-n)t^2+t^4}{(1-t^2)^3}.
\end{equation}
Under Laurent expansion around $s=0$, we obtain $a_0(\zeta)=a_1(\zeta)=(9-n)/8$. Notice that the singularities are toric for $n=0,\dots,3$. Therefore, these four rings are all stable as the symmetries are already maximal, and we will now only focus on $n\geq4$.

\paragraph{Case 1: dP$_4$} The PL of HS reads
\begin{equation}
\text{PL(HS)}=6t^2-5t^4+5t^6-\dots.
\end{equation}
There are 6 generators satifying 5 relations which can be written as \cite{2010arXiv1009.4044G}
\begin{eqnarray}
&&x_2x_6-x_3x_5+x_4^2,~x_2x_5-x_3x_4-x_6^2,~x_1x_6+x_2x_4-x_3^2-2x_5x_6,\nonumber\\
&&x_1x_5-x_2x_3+x_4x_6-2x_5^2,~x_1x_4-x_2^2+x_3x_6-2x_4x_5.\label{dP4eqn}
\end{eqnarray}
It turns out that the Gr\"obner basis consists of 6 equations:
\begin{eqnarray}
	&&x_4^2x_5-x_3x_5^2+x_3x_4x_6+x_6^3,~x_4^2-x_3x_5+x_2x_6,~x_1x_4-x_2^2-2x_4x_5+x_3x_6,\nonumber\\
	&&x_2x_4+x_1x_6-x_3^2-2x_5x_6,~x_1x_5-x_2x_3-2x_5^2+x_4x_6,~x_2x_5-x_3x_4-x_6^2.\label{dP4gb}
\end{eqnarray}
Let us first consider $\eta$'s that can be rescaled to some $\eta'$ that simultaneously make the initial terms to have $t$-weight zero. Then by \eqref{Futaki4},
\begin{equation}
\frac{8}{5}F=(v_1+v_2+v_3+v_4+v_5+v_6)\frac{3\times2-2}{2\times2}=v_1+v_2+v_3+v_4+v_5+v_6\label{dP4Fut}
\end{equation}
(where we have put $a_0$ on the left hand side). From the Gr\"obner basis, we see that there are monomials of various powers solely containing one $x_i$ without mixing for all $i\neq1$, so we only need to consider whether there is a test symmetry with charges $(-1,\dots)$ that destabilizes the ring in terms of the rescaling method. However, it has to be compensated by positive charges from more generators in the 6 equations as there are several mixing terms of form $x_1^px_{i\neq1}^q$ and they all have $p=q=1$. Alternatively, as it is sufficient to find one instance giving negative $F$ to destabilize the ring, we can also solve a system of inequalities: $2v_4+v_5\geq0,~v_3+2v_5\geq0,\dots$, together with $F\leq0$. It turns out there is no solution except $x_i=0$ to the inequalities\footnote{Notice this is a necessary but not sufficient condition for all the initial terms having a vanishing $t$-weight, but as it has no solutions, this certainly shows that there is no such $\eta'$ destabilizing the ring.}.

As an example, for the test symmetry $\eta$ with charges $(0,-1,-1,-1,-1,-1)$ (which would certainly lead to negative Futaki invariant if we do not modify the numerator or rescale it), the central fibre is
\begin{eqnarray}
&&x_4^2x_5-x_3x_5^2+x_3x_4x_6+x_6^3,~x_4^2-x_3x_5+x_2x_6,~-x_2^2-2x_4x_5+x_3x_6,\nonumber\\
&&x_2x_4+x_1x_6-x_3^2-2x_5x_6,~-x_2x_3-2x_5^2+x_4x_6,~x_2x_5-x_3x_4-x_6^2.
\end{eqnarray}
Consider $(\zeta+\epsilon\eta)$ as a Reeb field for small $\epsilon$, then the HS for \eqref{dP4eqn} (or equivalently \eqref{dP4gb}) is
\begin{equation}
	\text{HS}=\frac{1-5t^{4-2\epsilon}+5t^{6-3\epsilon}-t^{10-5\epsilon}}{\left(1-t^{2-\epsilon}\right)^5\left(1-t^2\right)}.
\end{equation}
We find $a_0(\zeta)=a_1(\zeta)=\frac{5}{2(\epsilon-2)^2}$. Thus, in our convention,
\begin{equation}
	F=n\text{D}_\epsilon a_1(\zeta+\epsilon\eta)-(n-1)\text{D}_\epsilon a_0(\zeta+\epsilon\eta)|_{\epsilon=0}=\frac{5}{8},
\end{equation}
which is positive as expected. More importantly, if we consider the test symmetry with $(1,0,0,0,0)$, this is the rescaled $\eta'$ we get from the above $\eta$ with equivalent test configuration. It does not receive any modifications in the numerator. Hence, we can use \eqref{dP4Fut} to compute the Futaki invariant, and indeed we get the same result $F=5/8$.

For $\eta$'s that cannot give zero $t$-weights to all the initial terms after rescaling, it is exhaustive to check all the cases. However, according to \cite{Kstabilitynotes}, we expect this ring to be stable.

\paragraph{Case 2: dP$_5$} The PL of HS reads
\begin{equation}
\text{PL(HS)}=5t^2-2t^4 \ ;
\end{equation}
the termination of the PL says that dP$_5$ is a complete intersection and it indeed is: the base Fano surface is a well-known degree 4 double-quadric in $\mathbb{P}^4$.
There are 5 generators satifying 2 relations which following theorem 115 in \cite{kaplan2013rational} can be written as
\begin{equation}
\sum_{i=1}^5x_i^2=\sum_{i=1}^5a_ix_i^2=0
\end{equation}
in $\mathbb{P}^4_{\mathbb{C}}$, where $a_i\neq a_j$ for $i\neq j$ and the subscript ``$\mathbb{C}$'' is explicit here just to emphasize that the field is algebraically closed as required by the theorem. By \eqref{Futaki4},
\begin{equation}
2F=(v_1+v_2+v_3+v_4+v_5)\frac{3\times2-2}{2\times2}=v_1+v_2+v_3+v_4+v_5.
\end{equation}
It suffices to check the test symmetry with charges $(1,0,0,0,0)$ due to the fact that all generators are symmetric within the relation. This symmetry indeed gives $F>0$. Hence, the ring is stable for $n=5$.

\paragraph{Case 3: dP$_6$} The PL of HS reads
\begin{equation}
	\text{PL(HS)}=4t^2-t^6.
\end{equation}
Again, this is a complete intersection: it is famous cubic surface in $\mathbb{P}^3$ with the 27 lines (in the PL, we have $-t^6$ because the generators are weighted by 2).
There are 4 generators satisfying 1 relation which can be written as
\begin{equation}
	x_1^3+x_2^3+x_3^3+x_4^3=0.
\end{equation}
By \eqref{Futaki4},
\begin{equation}
	\frac{8}{3}F=(v_1+v_2+v_3+v_4)\frac{3\times2-2}{2\times2}=v_1+v_2+v_3+v_4.
\end{equation}
It suffices to check the test symmetry with charges $(1,0,0,0,0)$ due to the fact that all generators are symmetric within the relation. This symmetry indeed gives $F>0$. Hence, the ring is stable for $n=6$.

\paragraph{Case 4: dP$_7$} The PL of HS reads
\begin{equation}
\text{PL(HS)}=3t^2+t^4-t^8.
\end{equation}
There are 4 generators satifying 1 relation which can be written as
\begin{equation}
x_1^4+x_2^4+x_3^4+x_4^2=0.
\end{equation}
By \eqref{Futaki4},
\begin{equation}
4F=(v_1+v_2+v_3)\frac{3\times2-2}{2\times2}+v_4\frac{3\times4-2}{2\times4}=v_1+v_2+v_3+\frac{5}{4}v_4.
\end{equation}
It suffices to check the test symmetries with charges $(1,0,0,0)$ and $(0,0,0,1)$ which both give $F>0$. Note here the generators $x_1$, $x_2$ and $x_3$ are symmetric in the relation. Hence, the ring is stable for $n=7$.

\paragraph{Case 5: dP$_8$} The PL of HS reads
\begin{equation}
\text{PL(HS)}=2t^2+t^4+t^6-t^{12}.
\end{equation}
There are 4 generators satifying 1 relation which can be written as
\begin{equation}
x_1^6+x_2^6+x_3^3+x_4^2=0.
\end{equation}
By \eqref{Futaki4},
\begin{equation}
8F=(v_1+v_2)\frac{3\times2-2}{2\times2}+v_3\frac{3\times4-2}{2\times4}+v_4\frac{3\times6-2}{2\times6}=v_1+v_2+\frac{5}{4}v_3+\frac{4}{3}v_4.
\end{equation}
It suffices to check the test symmetries with charges $(1,0,0,0)$, $(0,0,1,0)$ and $(0,0,0,1)$ which all give $F>0$. Note here the generators $x_1$ and $x_2$ are symmetric in the relation. Hence, the ring is stable for $n=8$.

As we can see, not all of the dPs are non-complete intersections (in fact, only dP$_4$ is a non-complete intersection). For instance, dP$_8$ is a complete intersection with
\begin{equation}
	\text{HS}=\frac{1-t^{12}}{\left(1-t^2\right)^2\left(1-t^4\right)\left(1-t^6\right)}.\label{dp8complete}
\end{equation}
Therefore, we can also use the standard steps for complete intersections to compute the Futaki invariant. One may check that this yields the same result as above. In fact, when writing the HS for test symmetry using PL, this recovers to the HS from complete intersection relation. Indeed, the degrees of the generators in PL(HS)$=2t^2+t^4+t^6-t^{12}$ agree with those in \eqref{dp8complete}. For instance, when we pick the test symmetry with non-vanishing charge on the generator at order 4, the HS becomes
\begin{equation}
	\text{HS}=\frac{\left(1-t^2+t^4\right)\left(1-t^4\right)}{\left(1-t^2\right)\left(1-t^2\right)^2\left(1-t^{4+\epsilon}\right)}.
\end{equation}
In particular,
\begin{equation}
	\frac{\left(1-t^2+t^4\right)\left(1-t^4\right)}{\left(1-t^2\right)}=\frac{1-t^{12}}{1-t^6}.
\end{equation}
Hence, we recover the HS in \eqref{dp8complete} with an explicit $1/(1-t^4)$ factor. As a result, the method for non-complete intersections is consistent with the method for complete intersections.
Importantly, our method is general and applies to arbitrary varieties.

\subsection{One SU($N$) Instanton Moduli Spaces on $\mathbb{C}^2$}\label{instaton}
The Higgs branch of D$(p-4)$-D$p$ brane systems, which is the moduli space of instantons, is studied in \cite{Benvenuti:2010pq}. Here, we consider the worldvolume theory of a D3 brane in the background of stack of $N$ D7 branes, whose $\mathcal{N}=1$ quiver is given in Figure 7 (with $k=1$) of \cite{Benvenuti:2010pq}. The U(1) factor of the global U($N$) global symmetry is absorbed into the gauge group U(1) in the quiver diagram. The superpotential is $W=q\Phi\tilde{q}$, where $q$ and $\tilde{q}$ are the fundamentals and $\Phi$ is a U(1) adjoint. Notice that there are two other U(1) adjoints $\phi_1$ and $\phi_2$ with superpotential term $\epsilon^{\alpha\beta}\phi_\alpha\Phi\phi_\beta$, but since the adjoints are just complex numbers for U(1), it vanishes in the superpotential. The HS is\footnote{Again, the fractional powers are always just computationally a result of our convention.}
\begin{equation}
	\text{HS}=\frac{\sum\limits_{i=0}^{N-1}\binom{N-1}{i}t^{2i/N}}{\left(1-t^{1/N}\right)^2\left(1-t^{2/N}\right)^{2(N-1)}}.
\end{equation}
The dimension of the moduli space is $n=2N$. Let us first consider the case with $N=2$. Under Laurent expansion around $s=0$, we have $a_0=a_1=8$. The PL of HS reads
\begin{equation}
	\text{PL(HS)}=2t^{1/2}+3t-t^2.
\end{equation}
Algebro-geometrically, we can write the equation as
\begin{equation}
	x_1^2+x_2^2+x_3^2+x_4^4+x_5^4=0.\label{oneSU2}
\end{equation}
If we consider the test symmetry with charges $(0,0,0,0,1)$, then we find that
\begin{equation}
	F=8\times\frac{4\times1/2-2}{2\times1/2}=0
\end{equation}
and
\begin{equation}
	||\eta||^2=\frac{(4-1)\times8}{4^2\times(4+1)\times(1/2)^2}=\frac{6}{5}\neq0.
\end{equation}
Hence, the ring is unstable. Interestingly, we can see that the central fibre is $x_1^2+x_2^2+x_3^2+x_4^4=0$, which is also known to be unstable from \S\ref{ADE}. Therefore, the destabilizing ring in general may not necessarily be stable as well.

If we further destabilize this $A_3$ threefold singularity with $(0,0,0,1,0)$, we would get the stable\footnote{Equivalently, we can consider $(0,0,0,1,1)$ for \eqref{oneSU2} to directly get this central fibre.} $x_1^2+x_2^2+x_3^2=0$, which is $\mathbb{C}^2/\mathbb{Z}_2$. In fact, if we remove the two $\phi_\alpha$'s in the quiver diagram, we would get the same superpotential and
\begin{equation}
	\text{HS}=\frac{1-t^2}{\left(1-t\right)^3},
\end{equation}
which in the IR fixed point should be the same as SQED with 2 flavours \cite{Aharony:1997bx}.

For general $N$, the varieties are not complete intersections. Even if we do not write the relations explicitly, we can still consider the test symmetry where only one generator of order $1/N$ has a charge 1 with other test charges vanishing. The Futaki invariant is then
\begin{equation}
	\frac{F}{a_0}=\frac{2N\times1/N-2}{2\times1/N}=0
\end{equation}
with
\begin{equation}
	\frac{||\eta||^2}{a_0}=\frac{2N-1}{(2N)^2\times(2N+1)\times(1/N)^2}=\frac{2N-1}{4(2N+1)}\neq0.
\end{equation}
Hence, the rings for one SU($N$) instanton moduli spaces are (K-)unstable.

\subsection{Phenomenological Theories}\label{pheno}
Now, let us consider the VMS of some phenomenologically interesting SUSY gauge theories.
\subsubsection{SQCD}\label{sqcd}
We can use the HS obtained in \cite{Gray:2008yu} to study the ring stabilities for SQCDs with SU($N_c$) gauge groups. The generators follow the standard relations between mesons and baryons: $B^{i_1\dots i_{N_c}}\tilde{B}_{j_1\dots j_{N_c}}=M^{[i_1}_{j_1}\dots M^{i_{N_c}]}_{j_{N_c}}$ and $M^{[i_1}_jB^{ji_2\dots i_{N_c}]}_{\mathstrut}=M^j_{[i_1}\tilde{B}^{\mathstrut}_{ji_2\dots i_{N_c}]}=0$.

\paragraph{Example 0: $N_f<N_c$} In such cases, the moduli spaces are freely generated, and the moduli spaces are simply $\mathbb{C}^{N_f^2}$ \cite{Gray:2008yu}. Hence, the HS is
\begin{equation}
	\text{HS}=\frac{1}{\left(1-t^{2/{N_f^2}}\right)^{N_f^2}}.
\end{equation}
As aforementioned, there are no non-trivial test configurations for $\mathbb{C}^{N_f^2}$. Hence, the rings for $N_f<N_c$ are stable. Notice, however, the discussion here is semi-classical. When we take quantum corrections into account, there is no stable\footnote{Here, this ``stable'' should not be confused with ``K-stable''.} ground state, and such vacuum variety is just an auxiliary space that helps us study the GIOs. For more details, see, for example, \cite{Gray:2008yu,Affleck:1984xz}.

\paragraph{Example 1: $N_f=2$, $N_c=2$} For $N_c=2$, the refined HS is
\begin{equation}
	\text{HS}=\sum_{k=0}^\infty\dim[0,k,0,\dots,0]t^{k/N_f}=~_2F_1\left(2N_f-1,2N_f;2;t^{1/N_f}\right),
\end{equation}
where $[n_1,\dots,n_{N_f-1}]$ is the highest weight notation of SU($N_f$) irrep, and $_2F_1$ is the hypergeometric function. In particular, for SU(2) gauge group, since the fundamentals are pseudoreal, there is no distinction between quarks and antiquarks. Moreover, as the fundamentals only have two colour indices, the antisymmetrized product on three or more flavour indices vanish. Hence, the relation becomes $\epsilon_{i_1\dots i_{2N_f}}M^{i_1i_2}M^{i_3i_4}=0$, where $i_1,\dots,i_{2N_f} = 1,\dots,2N_f$.

Let us start with $SU(2)$ with 2 flavours. The (unrefined) HS is
\begin{equation}
	\text{HS}=\frac{1-t}{\left(1-t^{1/2}\right)^6}.
\end{equation}
Under Laurent expansion around $s=0$, we have $a_0=a_1=64$. The PL of HS reads
\begin{equation}
	\text{PL(HS)}=6t^{1/2}-t,
\end{equation}
which is in fact a hypersurface. The defining equation is $x_1x_2+x_3x_4+x_5x_6=0$, or under a holomorphic change of coordinates, $u^2+v^2+w^2+x^2+y^2+z^2=0$. By \eqref{Futaki4},
\begin{equation}
	F=\sum_{i=1}^6v_i\frac{5\times1/2-2}{2\times1/2}a_0=32\sum_{i=1}^6v_i.
\end{equation}
It suffices to check test symmetry with charges (1,0,0,0,0,0) due to the symmetry of generators in the relation. We then have $F>0$. Hence, we conclude that the ring for SU(2) with $N_f=2$ is stable.

\paragraph{Example 2: $N_f=3$, $N_c=3$} The HS for SU(3) with 3 flavours is
\begin{equation}
\text{HS}=\frac{1-t^{2/3}}{\left(1-t^{1/3}\right)^2\left(1-t^{2/9}\right)^9}.
\end{equation}
Under Laurent expansion around $s=0$, we have $a_0=a_1=1162261467/256$. The PL of HS reads
\begin{equation}
	\text{PL(HS)}=9t^{2/9}+2t^{1/3}-t^{2/3}.
\end{equation}
There are 11 generators satisfying 1 relation which can be written as
\begin{equation}
	x_{11}x_{22}x_{33}+x_{21}x_{12}x_{33}+x_{11}x_{32}x_{23}+x_{21}x_{32}x_{13}+x_{31}x_{22}x_{13}+x_{31}x_{12}x_{23}+y_1y_2=0.
\end{equation}
By \eqref{Futaki4},
\begin{equation}
	F=\frac{1162261467}{256}\left(\frac{1}{2}(v_1+\dots+v_9)+2(v_{10}+v_{11})\right).
\end{equation}
As the mesons and baryons are symmetric in the single equation respectively and there are no mixing terms of mesons and baryons, the ring for SU(3) with 3 flavours is expected to be stable.

\paragraph{A speculation for $N_f=N_c$} More generally, as observed in \cite{Gray:2008yu}, the moduli space of $N_f=N_c$ is a hypersurface in $\mathbb{C}^{N_c^2+2}$ with
\begin{equation}
	\text{HS}=\frac{1-t^{2/N_c}}{\left(1-t^{2/N_c^2}\right)^{N_c^2}\left(1-t^{1/N_c}\right)^2}.
\end{equation}
Since a hypersurface can always have the initial terms with $t^0$ under rescaling, we can apply \eqref{Futaki4} which yields
\begin{equation}
	\frac{F}{a_0}=\frac{1}{2}(v_1+\dots+v_{N_c^2})+\frac{(N_c-1)^2}{2}(w_1+w_2)
\end{equation}
for test symmetry with charges $(v_1,\dots,v_{N_c^2},w_1,w_2)$. In particular, we have $F/a_0=1/2$ and $F/a_0=(N_c-1)^2/2$ for $(1,0,0,\dots,0)$ and $(0,0,\dots,0,1,0)$ respectively. The mesons and baryons are symmetric in the hypersurface algebraic equation with same $\zeta$-weights respectively, so in terms of the rescaling method it is natural to speculate that a negative $\eta$-charge of a generator would require other generators to have positive $\eta$-charges to compensate this in the test configuration. Moreover, there are no monomials having both mesons and baryons in the relation. Hence, it is natural to expect that the rings for $N_f=N_c$ are stable.

However, as we learn from \cite{Seiberg:1994bz} that the ring is expected to be (K-)unstable for $N_f<3N_c/2$. The (anti-)quarks have R-charges $(1-N_c/N_f)$, and therefore equal to zero for $N_f=N_c$. However, to have a conformal fixed point, we require the R-charges (of GIOs) to be no less than $2/3$, i.e., $N_f\geq3N_c/2$ here from the mesons. Thus, it seems that the K-stability criterion for conformality fails in this case.

\paragraph{Example 3: $N_f=4$, $N_c=3$} Even for non-zero R-charges, violation of unitarity bound might also ``leak'' out of the bound $nk-2<0$ from stability. For instance, the HS for SU(3) with 4 flavours reads
\begin{equation}
	\text{HS}=\frac{P(t)}{\left(1-t^{1/6}\right)^{16}\left(1-t^{1/4}\right)^2},
\end{equation}
where $P(t)$ is polynomial with palindromic coefficients whose exact expression can be found in \cite{Gray:2008yu} (up to some rescaling of $t$). Under Laurent expression, we learn that $n=16$. In fact, we can see that $N_f=4<3N_c/2=9/2$, and hence the mesons violate the unitarity bound. On the other hand, we have $nk-2=16/6-2=2/3>0$ for the mesons. Therefore, the unitarity bound could live above the stability bound.

\subsubsection{Electro-Weak MSSM}\label{ewsm}
The electroweak sectors of minimal supersymmetric standard model (MSSM) with renormalizable superpotentials are classified in \cite{He:2015rzg}. The simplest case is generated by $LH$ and $H\bar{H}$ where $L$ stands for the lepton doublets and $H$, $\bar{H}$ stand for the up and down types of Higgs doublets. Notice that we have suppressed the indices and Levi-Civita symbols in the generators. It turns out that geometrically this is just $\mathbb{C}^4$, and hence is trivially stable.

The next simplest case is generated by $LLe$ and $L\bar{H}e$ where $e$ stands for the lepton singlet. From \cite{He:2015rzg}, the HS is
\begin{equation}
	\text{HS}=\frac{1+4t+t^2}{(1-t)^5}.
\end{equation}
Under Laurent expansion around $s=0$, we have $a_0=a_1=729/16$. The PL of HS reads
\begin{equation}
	\text{PL(HS)}=9t-9t^2+16t^3-\dots
\end{equation}
There are 9 generators satifying 9 relations which can be written as
\begin{eqnarray}
	&&y_{6}y_{8}-y_{5}y_{9},~y_{3}y_{8}-y_{2}y_{9},~y_{6}y_{7}-y_{4}y_{9},\nonumber\\ &&y_{5}y_{7}-y_{4}y_{8},~y_{3}y_{7}-y_{1}y_{9},~y_{2}y_{7}-y_{1}y_{8},\nonumber\\ &&y_{3}y_{5}-y_{2}y_{6},~y_{3}y_{4}-y_{1}y_{6},~y_{2}y_{4}-y_{1}y_{5},
\end{eqnarray}
which already forms a G\"obner basis. For those $(v_1,v_2,\dots,v_9)$ that can be rescaled such that all the 9 equations have initial terms with 0 $t$-weights, we can simply apply \eqref{Futaki4} which yields
\begin{equation}
	F=\frac{729}{16}\times\frac{3}{2}\sum_{i=1}^9v_i=\frac{2187}{38}\sum_{i=1}^9v_i.\label{y1-9}
\end{equation}
Due to the symmetry of the 9 variables, if there is a negative test charge, then it should be compensated by more positive test charges in order to satisfy the condition for a rescaled configuration. Hence, \eqref{y1-9} should always give a positive $F$.

However, for the test symmetries that cannot be rescaled to one where \eqref{y1-9} applies, it is exhaustive to check all of them. As an example, let us consider $\eta$ with charges $(-1,-2,0,0,\dots,0)$. The test configuration is then
\begin{eqnarray}
&&y_{6}y_{8}-y_{5}y_{9},~y_{3}y_{8}-t^{-2}y_{2}y_{9},~y_{6}y_{7}-y_{4}y_{9},\nonumber\\ &&y_{5}y_{7}-y_{4}y_{8},~y_{3}y_{7}-t^{-1}y_{1}y_{9},~t^{-2}y_{2}y_{7}-t^{-1}y_{1}y_{8},\nonumber\\ &&y_{3}y_{5}-t^{-2}y_{2}y_{6},~y_{3}y_{4}-t^{-1}y_{1}y_{6},~t^{-2}y_{2}y_{4}-t^{-1}y_{1}y_{5}.
\end{eqnarray}
With the help of \texttt{Macaulay2}, a direct computation with regularization in the numerator yields
\begin{eqnarray}
	\text{HS}&=&\frac{1}{(1-t)^7 \left(1-t^{1-2 \epsilon }\right)
		\left(1-t^{1-\epsilon }\right)}\times\left(1-3t^2-4 t^{2-2 \epsilon }-2 t^{2-\epsilon }\right.\nonumber\\
	&&\left.+2 t^{3-3 \epsilon }+9 t^{3-2 \epsilon }+3 t^{3-\epsilon }+2 t^3-3 t^{4-3 \epsilon }-6 t^{4-2 \epsilon }+t^{5-2 \epsilon }-t^{5-\epsilon }+t^{6-3 \epsilon }\right).\nonumber\\
\end{eqnarray}
Thus,
\begin{equation}
	F=n\text{D}_\epsilon a_1(\zeta+\epsilon\eta)-(n-1)\text{D}_\epsilon a_0(\zeta+\epsilon\eta)|_{\epsilon=0}=\frac{1}{2}.
\end{equation}
We also notice that this $\eta$ can be rescaled to the ``minimal'' $\eta'$ with charges $(1,0,2,0,0,\dots,0)$. The test configuration is then
\begin{eqnarray}
&&y_{6}y_{8}-y_{5}y_{9},~t^2y_{3}y_{8}-y_{2}y_{9},~y_{6}y_{7}-y_{4}y_{9},\nonumber\\ &&y_{5}y_{7}-y_{4}y_{8},~t^2y_{3}y_{7}-ty_{1}y_{9},~y_{2}y_{7}-ty_{1}y_{8},\nonumber\\ &&t^2y_{3}y_{5}-y_{2}y_{6},~t^2y_{3}y_{4}-ty_{1}y_{6},~y_{2}y_{4}-ty_{1}y_{5}.
\end{eqnarray}
Regularization in the numerator yields
\begin{equation}
	\text{HS}=\frac{1-2 t^{2+\epsilon}-7 t^2+5 t^{3+\epsilon}+11 t^3-3 t^{4+\epsilon}-6 t^4-t^{5+\epsilon}+t^5+t^{6+\epsilon}}{(1-t)^7 \left(1-t^{1+\epsilon}\right) \left(1-t^{1+2 \epsilon}\right)}.
\end{equation}
Therefore, we find that
\begin{equation}
F=n\text{D}_\epsilon a_1(\zeta+\epsilon\eta)-(n-1)\text{D}_\epsilon a_0(\zeta+\epsilon\eta)|_{\epsilon=0}=\frac{1}{2}.
\end{equation}
We have checked quite a few test symmetries with low values of $v_i$, all of which give positive Futaki invariants. It is natural to speculate that this ring is stable.

\section{Conclusions and Outlook}\label{outlook}

\begin{table}
\begin{tabular}{c|c}\\ 
Affine Variety & K-Stability \\ \hline
Toric & all are stable; this is well-known \\
Type A 3-folds:  $w^2 + x^2 + y^2 + z^{n+1} = 0$ & stable for $0 < n < 3$ \\
Type D 3-folds:  $w^2 + x^2 + y^2z + z^{n} = 0$ & stable only for $ n = 3$ \\
Type E 3-folds & stable \\
Cone over del Pezzo surfaces & stable (for all 9 cases $n=0, \ldots, 8$) \\
One SU($N$) instanton moduli space on $\mathbb{C}^2$ & unstable \\
SQCD for $N_f = N_c$ & expected to be stable (checked $N_f = 2,3$) \\
(Simplest) Electro-weak MSSM & stable
\end{tabular}
\caption{
K-stability of some of the illustrative examples considered in this paper. The ADE threefolds were also systematically studied in \cite{2015arXiv151207213C,Fazzi:2019gvt}.
\label{t:sum}
}
\end{table}

In this paper, we studied the K-stability of chiral rings, and tested our results on several examples (for the reader's convenience, we summarize some of the key results in Table \ref{t:sum}). 
By considering the PL, we can apply the calculations in \cite{Collins:2012dh,2015arXiv151207213C,Collins:2016icw} to general varieties, and non-complete intersections in particular. 
We found that when considering a test symmetry, it may not be enough to only incorporate $\epsilon$'s to the denominators. 
In fact, we should write the HS with respect to $(\zeta+\epsilon\eta)$, which is the Reeb field for sufficiently small $\epsilon$. This is because $(\zeta+\epsilon\eta)$ is treated as a Reeb vector field in the derivation of Futaki invariant in \cite{Collins:2012dh}.  However, notice that the new R-symmetry we obtain is still $\zeta(\epsilon)=(1-a\epsilon)\zeta+\epsilon\eta$, which does not affect $(\zeta+\epsilon\eta)$ from being a Reeb field even if the minimum of $a_0(\zeta(\epsilon))$ is reached at some $\epsilon>0$.

When we write the HS with respect to $(\zeta+\epsilon\eta)$, we still start with the multi-graded (refined) HS where the small $\epsilon$ has not appeared. Therefore, it is still not homogenous with repect to $\eta$. We proposed that we should use the $t$-weighting induced by $\eta$ as an ordering for the initial terms to write the HS perturbed by $\epsilon$. We also saw that though this works very well, for some ``strange'' (trivial) test symmetries with $F=0$ such as $(1,-1,0,0)$ for $uv=xy$, $c_0-b_0^2/a_0$ is not zero and we still need unusual definition of the norm. We found that by including higher corrections of $\epsilon$ in the numerator, $c_0-b_0^2/a_0$ would become zero. However, this could possibly be a coincidence, and it would be interesting to further study this problem.

For arbitrary rings, there is still not a clear way to reduce the number of test symmetries one need to consider. This would be very crucial when we have more variables and relations. For instance, we have not discussed SQCD with $N_f>N_c$, more complicated geometries of electroweak MSSM or that of the entire MSSM (whose HS was computed in \cite{Xiao:2019uhh}). 
It is computationally hard to go through all the test symmetries and we proposed a rescaling method, so that the calculations could be more or less simplified. However, more details and evidences for this still need to be explored. We argued that if this works, then there are only two possibilies to destabilize a ring. Either there is a small enough $\zeta$-weight $k$ such that $nk-2<0$, or there is a generator with negative test charge which is cancelled in the equations (so that no monomial would have $t^p$ with $p<0$) but it has a large enough $k$ which makes the Futaki invariant negative. These different ways of destabilizing the ring might probably be related to different physical interpretations. We need to have a deeper understanding of the physics behind the destabilizing process, and it might also in turn be helpful to determine what test symmetries should be considered.

K-stability is naturally related to the chiral rings of SCFTs as some ``generalized $a$-maximization''. However, when an AdS/CFT picture is not present, the connection between K-stability and conformality becomes more subtle. However, as an example, we show that SQCD does not seem to follow the K-stability criterion for conformality. Furthermore, the unitarity bound is possible to live above the stability bound $nk-2\geq0$, so some operators which violate the unitarity bound could have positive $nk-2$. Nevertheless, K-stability should still play a crucial role in studying chiral rings and SCFTs since on the (emergent) gravity side, there usually involves many symmetries, and this is exactly what K-stability and destabilizing rings concern. We speculate that K-stability could be a \emph{necessary} (but not sufficient) condition for the ring being a ring of SCFT. This condition might become sufficient as well in some special classes of theories, such as the gauge theories from D-branes probing CYs.

In \cite{Benvenuti:2017lle}, chiral ring stability is introduced when one drops certain superpotential terms. Its relation to K-stability still requires further study. It is also worth noting that in \cite{Fazzi:2019gvt}, non-commutative crepant resolution (NCCR) is applied to finding the quivers for various theories. However, the existence of NCCR and being K-stable are not necessary to each other. It would be interesting to further study their connections and also extend the discussions to supersymmetric theories in other dimensions.

\section*{Acknowledgements}
We are grateful to Tristan Collins, G\'abor Sz\'ekelyhidi and Alessandro Tomasiello for enlightening discussions. We would also like to thank Natthawut Phanachet for initial collaborations. 
YHH is indebted to STFC for grant ST/J00037X/1.

\appendix

\section{Gr\"obner Bases \& Hilbert Series}\label{gb}
Since our chiral rings can be realized as quotient rings of polynomial rings over $\mathbb{C}$ by defining ideals arising from the likes of polynomial F-terms, it is important for us to systematically study such objects. The first step toward any serious investigation of an ideal $I$ within a graded ring is the establishment of its Gr\"obner basis GB($I$); constituting the pillar of computational algebraic geometry \cite{schenck_2003,M2} (cf. \cite{compbook} for recent advances and applications in the context of gauge/string theories).

Briefly \cite{schenck_2003,sturmfels1991}, for the polynomial ring $R=\mathbb{C}[x_1,x_2,\dots,x_n]$ to any monomial $\vec{x}^{\vec{\alpha}}:=x_1^{\alpha_1}x_2^{\alpha_2}\dots x_n^{\alpha_n}$ with each $\alpha_i \in \mathbb{Z}_{\geq 0}$ (the short-hand notation of raising the exponent is standard) in $R$, we can associate the exponent vector $\vec{\alpha}$; this defines a monomial ordering $\succ$ such that
\begin{enumerate}
	\item $\succ$ is a total order on $R$, i.e., for any elements $\vec{\alpha}, \vec{\beta}$, one and only one of the three possibilities 
	$\vec{\alpha} \succ \vec{\beta}$, or $\vec{\beta} \succ  \vec{\alpha}$, or $ \vec{\alpha} = \vec{\beta}$ occurs;
	\item for any $\vec{\gamma}$, if $\vec{\alpha} \succ\vec{\beta}$, then $\vec{\alpha} + \vec{\gamma} \succ \vec{\beta} + \vec{\gamma}$;
	\item  $\succ$ is a well-ordering in that any nonempty subset has a smallest element.
\end{enumerate}
Of course, these properties are no more than the axiomatization of how we usually manipulate degrees in monomials. Indeed, we will denote total degree of a monomial as $|\vec{\alpha}| = \sum\limits_{i=1}^n \alpha_i$.

We emphasize that there are many possible choices of this ordering and the most typical are
\begin{itemize}
	\item {\sf Lexicographic: } this is just dictionary ordering, i.e., $\vec{\alpha} \succ_{{\rm Lex}} \vec{\beta}$ if the leftmost nonzero entry of $\vec{\alpha} - \vec{\beta}$ is positive;
	\item {\sf Graded Lexicographic: } this is sorting by total degree first and then by lexicographic, i.e., $\vec{\alpha} \succ_{{\rm grLex}} \vec{\beta}$ if $|\vec{\alpha}| > |\vec{\beta}|$ or, when $|\vec{\alpha}| = |\vec{\beta}|$, we have $\vec{\alpha} \succ_{{\rm Lex}} \vec{\beta}$. There is a reverse version of this where one sorts by total degree first and then if they are equal, then $\vec{\alpha} \succ_{{\rm grevLex}} \vec{\beta}$ if the rightmost nonzero entry of $\vec{\alpha} - \vec{\beta}$ is negative;
	\item {\sf General Weighted Lexicographic: } We can weight each variable $x_i$.
	For example, choose a weight vector $\vec{w} = (w_1, w_2, \ldots, w_n)$ for the variables $x_i$. Usually, the weight is taken to be $w_i \in \mathbb{Z}_{\ge 0}$.
	This weight can, for example, be prescribed by the R-charges.
	Here, the total degree is obviously $|\vec{\alpha}| = w \cdot \vec{\alpha}$.
	
	In fact, one is not restricted to just weighting each variable by some non-negative integer but in general by some vector, say of length $k \le n$, so that we have some {\it weight matrix} $W_{k \times n}$. Then we could sort as: $\vec{\alpha} \succ_W \vec{\beta}$ if $W \cdot \vec{\alpha} \succ_{{\rm Lex}} W \cdot \vec{\beta}$. This multi-weighting can be used as a {\it refinement} of possible charges and variables thus graded are called fugacities \cite{Feng:2007ur,Forcella:2008bb}. 
\end{itemize}

An example, taken from \cite{schenck_2003}, would illustrate the above. Suppose $R=\mathbb{C}[x,y,z]$, and we weight $x,y,z$ with the standard base vectors $(1,0,0)$, $(0,1,0)$ and $(0,0,1)$, then $x \succ_{{\rm Lex}} yz^2$ since $(1,0,0) - (0,1,2)$ has the leftmost entry $1$ which is positive. On the other hand, $yz^2 \succ_{{\rm grLex}} x$ since the degrees are $|x| = 1$ and $|yz^2| = 3$; this graded lexicographic ordering is one perhaps most familiar to us.

Having fixed a monomial ordering $\succ$ on $R$, then we have
\begin{definition}
	For any multivariate polynomial $f = \sum\limits_{\vec{\alpha}} c_{\vec{\alpha}} \vec{x}^{\vec{\alpha}} \in R$, the \emph{initial monomial} $\textup{in}(f)$ is the largest (with respective to $\succ$) monomial term in $f$. We can always make the coefficient of this term to be 1 so that $f$ is monic.
\end{definition}
Thus prepared, we are finally at the crux of our subject:
\begin{definition}
	A subset $\{g_1, g_2, \ldots, g_m \}$ for an ideal $I$ is a Gr\"obner basis $\textup{GB}(I)$ for $I$ if the ideal generated by the initial monomials of the elements of $I$ is generated by $\{\textup{in}(g_1), \ldots, \textup{in}(g_m)\}$, i.e., if
	\[
	\textup{in}(I) = \langle \textup{in}(g_i) \rangle .
	\]
\end{definition}

Computationally, we have the important result that
\begin{theorem}
	A set $G$ is a Gr\"obner basis iff the S-polynomial (or syzygy pair) defined as
	\[
	S(g_i, g_j) := \frac{\textup{lcm}\big( \textup{in}(g_i), \ \textup{in}(g_j)\big)}{\textup{in}(g_i)} g_i -
	\frac{\textup{lcm}\big( \textup{in}(g_i), \ \textup{in}(g_j)\big)}{\textup{in}(g_j)} g_j
	\]
	reduces modolo $G$ for \emph{all} pairs $g_i, g_j \in G$.
\end{theorem}
This gives a practical - albeit exponential-running-time - algorithm, the so-called {\it Buchberger algorithm} for computing GB($I$) given an ideal $I = \langle{f_i}\rangle_{i=1,\ldots,N}$:
\begin{enumerate}
	\item Set $G = \left\{ f_1, \ldots, f_N \right\}$ and compute $S(f_i,f_j)$ for each of the pairs with respect to a chosen ordering $\succ$;
	\item Compute the remainder of each $S(f_i,f_j)$ upon division by each of the elements of $G$. If the remainder is not zero, then include this $S(f_i,f_j)$ as a new element of $G$;
	\item Repeat until all remainders with respect to all elements are 0; this final list (which could have much more than $N$ elements) is a Gr\"obner basis for $I$.
\end{enumerate}

\subsection{Hilbert Series: Revisited}\label{hsrev}
In light of the discussions above, more properties, especially from a computational perspective, of the HS emerge. Most importantly, we have a the classical result of Macaulay \cite{BAYER199231} that
\begin{theorem}
	The Hilbert series of $\textup{in}(I)$ is the same as that of the ideal $I$ itself.
\end{theorem}
Thus explicit computation of the HS reduces to finding the Gr\"obner basis: given the ideal $I$, we simply (1) compute its Gr\"obner basis GB$(I) = \{g_i\}$ with respect to some monomial ordering; (2) find the initial ideal $\langle\text{in}(g_i)\rangle$ (this is a Gr\"obner basis guarantees that this ideal is equal to in($I$)); (3) importantly each generator in$(g_i)$ is monomial and we thus only need to compute the basis of monomials modolo these monomials at each degree and sum the generating series to obtain the HS for in$(I)$, which by the above theorem is then the HS for $I$.

Moreover, one can \emph{refine} the HS: this means we can assign not just a single weight to the variable $t$, but, instead, a vector of weights for multi-variables $t_i$. In other words, the polynomial ring will be multi-graded. For example, for $\mathbb{C}^3$, the (unrefined) HS is $\text{HS}(t;\mathbb{C}^3) = (1-t)^{-3}$ and the refined series can be, for instantce, $\text{HS}(t_1,t_2,t_3; \mathbb{C}^3) = \left((1-t_1)(1-t_2)(1-t_3)\right)^{-1}$.

\addcontentsline{toc}{section}{References}
\bibliographystyle{utphys}
\bibliography{references}

\end{document}